\documentclass[aip,reprint,floatfix]{revtex4-2}  %tightenlines>>> use for US letter paper
\usepackage[]{graphicx}
\usepackage{dcolumn}%
\usepackage{bm}%
\usepackage{amsmath}%
\let\baraccent=\= % rename builtin command \= to \baraccent
\renewcommand{\=}[1]{\stackrel{#1}{=}} % for putting nums above =

\begin{document} 
%\pagestyle{fancy}
%\title{Ultrabroadband photoconduction trap density trends in amorphous oxide semiconductors }
%\title{Capturing trends in trap density for amorphous oxide semiconductors captured using ultrabroadband photoconduction }
\title{Illuminating trap density trends in amorphous oxide semiconductors with ultrabroadband photoconduction }
\author{George W. Mattson$^1$, Kyle T. Vogt$^1$, John F. Wager$^2$ and Matt W. Graham$^1$}
\affiliation{Department of Physics, Oregon State University, Corvallis, OR 97331-6507, USA} 
\affiliation{School of EECS, Oregon State University, Corvallis, OR 97331-5501, USA}

\begin{abstract}
\indent	
\\

 Under varying growth and device processing conditions, ultrabroadband photoconduction (UBPC) reveals strongly evolving trends in the defect density of states (DoS) for amorphous oxide semiconductor thin-film transistors (TFTs).  Spanning the wide bandgap of amorphous InGaZnO$_x$ (a-IGZO), UBPC identifies seven oxygen deep donor vacancy peaks that are independently confirmed by energetically matching to photoluminescence emission peaks.   The sub-gap DoS from 15 different types of a-IGZO TFTs all yield similar DoS, except only back-channel etch TFTs can have a deep acceptor peak seen at 2.2 eV below the conduction band mobility edge.  This deep acceptor is likely a zinc vacancy, evidenced by trap density which becomes 5-6$\times$ larger when TFT wet-etch methods are employed. Certain DoS peaks are strongly enhanced for TFTs with active channel processing damage caused from plasma exposure.   While Ar implantation and He plasma processing damage are similar, Ar plasma yields more disorder showing a $\sim 2\times$ larger valence-band Urbach energy, and two orders of magnitude increase in the deep oxygen vacancy trap density. Changing the growth conditions of a-IGZO also impacts the DoS, with zinc-rich TFTs showing much poorer electrical performance compared to 1:1:1 molar ratio a-IGZO TFTs owing to the former having a $\sim$10$\times$ larger oxygen vacancy trap density. Finally, hydrogen is found to behave as a donor in amorphous indium tin gallium zinc oxide TFTs.  

\end{abstract}

\maketitle

\section{Introduction}
Amorphous oxide semiconductors (AOS) such as amorphous indium gallium zinc oxide (a-IGZO) have achieved widespread adoption as the active channel material in optical display thin-film transistors (TFTs) because of their high mobility, low processing cost, and high on-off current ratio.\cite{Kamiya2010, Wager2014} a-IGZO is a wide bandgap semiconductor material with $\mathrm{E_g}$ ranging from 3.1 to 3.5 eV. Owing to its amorphous structure, a-IGZO has a large concentration of sub-gap vacancy sites that serve as the dominant electron donation mechanism for its n-type TFT operation.\cite{Nomura2004, Kamiya2010} The subgap states in a-IGZO also act as electron traps that impact device performance by introducing transfer curve hysteresis and bias illumination stressing.\cite{Cho2011, Ahn2014, Vogt2020, Kim2015, Yoon2018, Zhang2019} Fabrication processes for AOS TFTs have a marked effect on the overall characteristics of the resulting devices, which is reflected in the composition of the subgap states.

Energetically, these trap states span different ranges of the subgap depending on their local charge environment and consequent trapping behavior. Multiple configurations of donor-like oxygen vacancies ($\mathrm{V_{O}}$) dominate the shallow and midgap regions.\cite{Fung2009, Chen2011, Song2016, Song2018} Additionally, both acceptor-like zinc vacancies ($\mathrm{V_{Zn}}$)\cite{Vogt2020} and an OH-related state ($\mathrm{{[{O_{O}^{2-}}{H^+}]}^{1-}}$)\cite{Mattson2022} have been documented; these states are convolved with an exponentially decaying Urbach\cite{Urbach1953} tail reflecting O 2p disorder in the amorphous matrix.\cite{Wager2017}

The donor-like $\mathrm{V_{O}}$ states have received considerable attention from researchers over the years.\cite{Fung2009, Chen2011, Song2016, Song2018}  $\mathrm{V_{O}}$ are the dominant donor-like trap state in AOS materials. In a-IGZO, this electron donation pins the Fermi energy ($\mathrm{E_F}$) near the CBM, making them n-type semiconductors. While electron donation from thermally depopulated shallow donor $\mathrm{V_{O}}$ states typically contributes the n-type carriers necessary for AOS enhancement-mode operation, $\mathrm{V_{O}}$ states have been variously linked to effects such as hysteresis\cite{deJamblinnedeMeux2017} and persistent photocurrents\cite{Wang2022}. Process tuning of $\mathrm{V_O}$ trap states to amplify AOS TFT electronic performance while avoiding deleterious stability or leakage phenomena remains an ongoing research challenge.

This work reveals trends in how the subgap density of states (DoS) in AOS TFTs evolves over different processing methods, TFT architectures, and channel compositions. In section III, we employ the ultrabroadband photoconduction (UBPC) method to  reveal systematic trends in subgap DoS.  Section III is broken down into five sub-sections that explore the DoS of AOS materials in the context of: (A) Comparison ultrabraodband optical vs. UBPC  photoconductive methods of defect state identification, (B) etch process-induced vacancy state formation in back-channel etch vs. top-gate a-IGZO TFTs, (C) plasma and ion implantation treated a-IGZO top-gate TFTs, (D) a-IGZO TFTs with non-stoichiometric active channel compositions, and (E) hydrogen incorporation into amorphous indium tin gallium zinc oxide (a-ITGZO).  Below, we summarize the motivation for each study.

Section III.A compares the subgap defect peaks identified by UBPC in AOS TFTs by matching peaks obtained to more conventional optical methods, including photoluminescence (PL). AOS TFT device electrical metrics have improved over the last two decades.  The optical response at subgap defect energies in AOS materials is often $\mathrm{\sim 10^{6}}$ smaller than at the bandgap. While  x-ray or ultraviolet photoelectron spectroscopy (XPS, UPS) have become popular tools for AOS characterization, they are not capable of achieving the necessary resolution to reliably observe the subgap trap states.\cite{Rajachidambaram2012, Du2014}   The recently developed UBPC method has the potential to be a standard experimental technique to quantify the subgap trap state density analytically. 

Section III.B  statistically compares the DoS trends for back channel etch (BCE) vs. top-gate (TG) a-IGZO TFTs. Fabrication of BCE AOS TFTs is convenient for TFT arrays in display backplane applications.\cite{Kim2007} However, multiple researchers have observed that the etch process traditionally employed in fabricating BCE a-IGZO TFTs results in device degradation,\cite{Kim2007} ostensibly due to the creation of metal vacancy states which are introduced via the etchant.\cite{Kwon2010, Nag2014, Park2015}

Section III.C compares the change in DoS of a-IGZO TFTs after  plasma and ion implantation methods are applied to the full active channel.  Extensive prior research has been performed on plasma\cite{Park2007, Kim2012plasma, Kim2014, Hwang2014, Liu2015, Lu2016, Jang2018, Um2018, Liu2021, Park2022plasma} and ion implantation\cite{Yasuta2021} treatment of a-IGZO TFTs.    Such plasma treatment have a variety of purposes  ranging from defect creation\cite{Park2007, Jang2018} to defect passivation\cite{Kim2012plasma, Ding2016, Abliz2020} to use as a dry etchant for the source-drain electrode metal.\cite{Joo2013, Wang2015}  In a-IGZO, Such plasma methods commonly increase the TFT conductivity up to $\mathrm{\sim 50X}$which is desirable for improving  electrical contacts.\cite{Park2007, Kim2014, Jang2018}  After plasma treatments, the UBPC method will be used to understand how oxygen vacancy deep donor trap density is correlated with conductivity enhancement.  The plasmas selected for these treatments vary greatly depending on the intended application: oxygen\cite{Kim2014, Ding2016} plasmas can passivate $\mathrm{V_O}$ defects, while other plasmas can dope the device CB via either hydrogen\cite{Kim2012plasma, Abliz2020, Park2022plasma} incorporation or the formation of $\mathrm{V_O}$ states (Ar, He\cite{Jang2018} plasmas); mixed Ar-O\cite{Hwang2014, Liu2015, Liu2021} plasmas have also been explored.

In section III.D,  AOS growth conditions are changed by comparing the DoS of a zinc-rich growth to that of  `111' (1:1:1 molar \% $\mathrm{In_{2}O_{3}/Ga_{2}O_{3}/ZnO}$ ratio) a-IGZO. Exploring non-stoichiometric compositions of a-IGZO (with greater or lesser relative concentrations of the In, Ga, and/or Zn metal constituents compared to 111 a-IGZO) is an active area of research for improving upon a-IGZO TFT characteristics such as field effect mobility.\cite{Kamiya2009, Kim2012, Hsu2015, Jeong2016, Hu2017, Koretomo2020} These efforts are sometimes linked to the deposition of bilayer\cite{Marrs2011, Jung2014, Tai2019, Cho2021, Choi2022} (or trilayer)\cite{Liu2017, Dargar2019, Liu2022} active channels. These bilayer TFTs often contain a stoichiometric or near-stoichiometric stability layer\cite{Jeong2016} with a boost layer that is richer in certain constituent elements such as In\cite{Koretomo2020, Cho2021}, doped with metals such as Ti\cite{Hsu2014}, or composed of a conductive material such as ITO\cite{Stewart2017}) intended to enhance the TFT electrical performance.

Lastly, Section III.E discusses hydrogen incorporation into indium tin gallium zinc oxide (a-ITGZO) TFTs. a-ITZGO has been studied in recent years to reduce the reliance of AOS TFTs on the scarce indium constituent without the adverse device mobility effects that might be expected from reducing or eliminating the molar proportion of indium.\cite{Lee2020, Kim2020, Kong2022} While a-ITGZO TFTs exhibit promising carrier mobility metrics, they also exhibit high susceptibility to positive bias stressing (PBS)-related degradation effects attributed to subgap trap states.\cite{Kim2020} 
Hydrogen can be an abundant defect in AOS materials \cite{Bang2017}, but the electrically active nature of this large interstitial defect is more controversial. Recent work suggests this state that results from its interaction with oxygen states located near the valence band mobility (VBM) edge.\cite{Mattson2022}  Although hydrogen has a donor effect on a-IGZO, its negative-U nature results in a hydrogen electron donation that precedes its incorporation into a-IGZO. \cite{VanDeWalle2003, VanDeWalle2006,Mattson2022}   Specifically for a-IGZO TFTs, the hydrogen complex state $\mathrm{{[{O_{O}^{2-}}{H^+}]}^{1-}}$ has been observed centered $\mathrm{\sim 0.4}$ eV below the VBM edge.\cite{Wager2017,Mattson2022}  Section III.E extends prior hydrogen studies to explore other AOS TFT materials.

\section{Experimental Methods}

\subsection{Amorphous oxide TFT device characterization}
Trends in the subgap DoS are observed for a diverse selection of AOS back channel etch (BCE) and top-gate a-IGZO TFTs discussed in each sub-section. ITO/a-IGZO co-sputtered (a-ITGZO) TFTs possessing different hydrogen concentrations are also fabricated. The TFTs possessing different hydrogen concentrations were synthesized by subjecting the devices to varying annealing time intervals to cause hydrogen migration from an H-rich $\mathrm{SiN_y}$ passivation layer adjacent to the active channel.\cite{Mattson2022}
\begin{figure}
   \begin{center}
   \includegraphics[height=6cm]{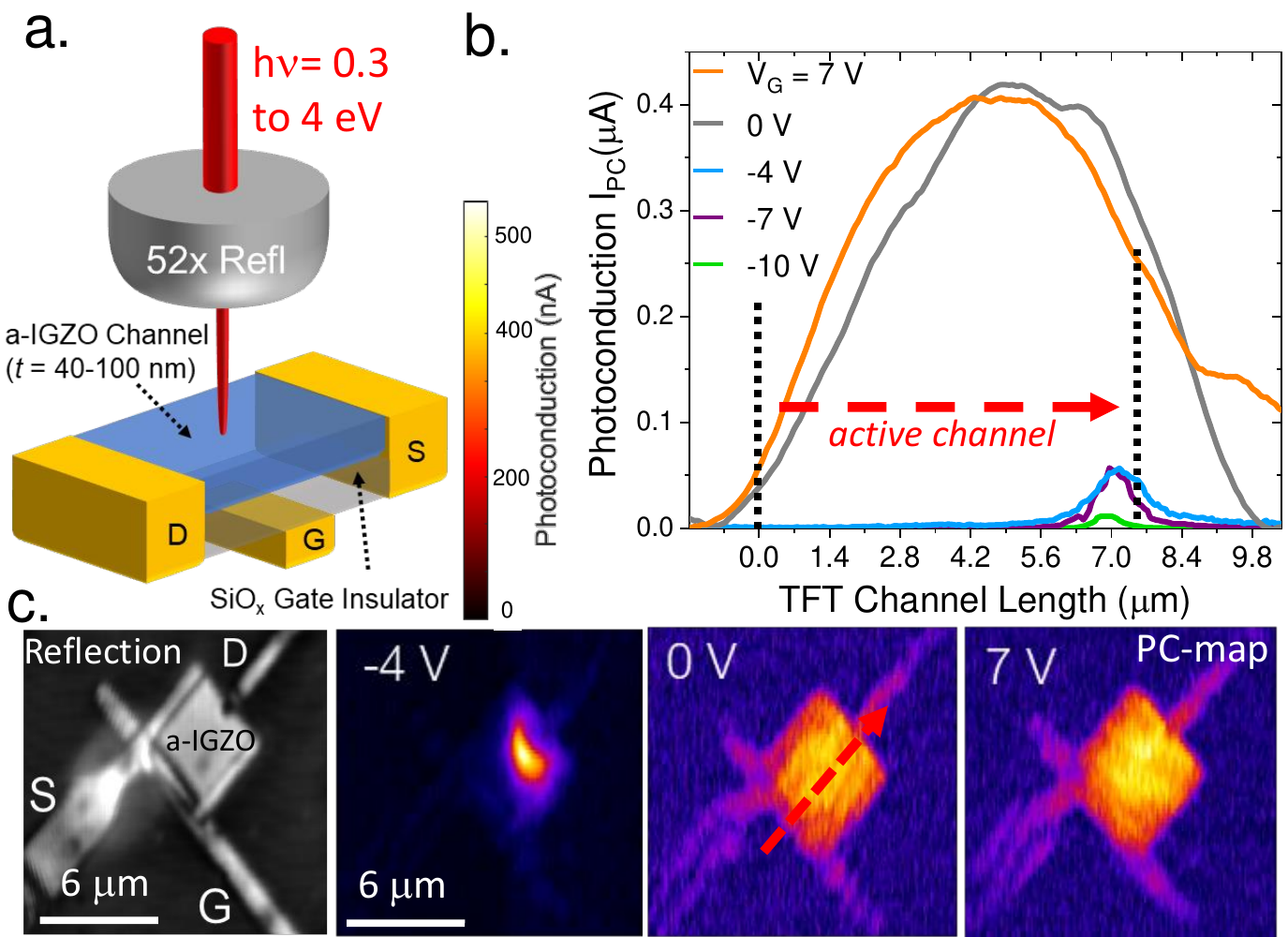}

   \end{center}
   \caption[example] 
%>>>> use \label inside caption to get Fig. number with \ref{}
   { \label{fig:fig5pt1} (\textbf{a}) The UBPC method illustrated uses tunable lasers and an all-reflective 4\textit{f} confocal scanning microscope to measure the DoS for the AOS TFTs to within 0.3 eV of the CB mobility edge.  (\textbf{b}) Shown for $h \nu= 1.5$ eV, UBPC microscopy spatially-resolves the TFT channel length PC-response, $\mathrm{I_{PC}}$. Gate voltages resolve 'turn-on' change from depletion-mode ($\mathrm{V_G < 0}$ V) to enhancement-mode ($\mathrm{V_G > 0}$ V) operation.  (\textbf{c}) Scanning-PC microscopy spatial maps corresponding to (\textit{red arrow}) linecuts shown in b with back-reflection map of a BCE a-IGZO TFT on right.} 
   \end{figure} 
The  photoluminescence (PL) emission spectra of a-IGZO thin films and devices were taken using a Horiba Nanolog flourimeter using using photomultiplier detection for the visible range, and liquid nitrogen-cooled InGaAs detection over the 0.7 to 1.5 eV range.  All PL data shown uses a 3.8 eV excitation source. PL spectral shape on active TFTs was confirmed using an Ocean Optics spectrometer under diffraction-limited illumination. The corresponding ultrabroadand absorption spectrum for Tauc bandgap analysis was taken using a Cary UV-VIS-IR spectrometer. 

\subsection{Sub-gap DoS by Ultrabroadband Photoconduction (UBPC) microscopy}
In this work, the on-chip spectroscopic technique called ultrabroadband photocoduction (UBPC)\cite{Vogt2020} is employed to obtain the experimental subgap trap density and DoS for AOS TFTs. The essential elements of the UBPC setup are depicted in Fig. 1(a). A laser source tunable over the a-IGZO subgap energy range is focused onto the TFT active channel using a piezo scanning mirror within a 4-f confocal scanning geometry which couples the source into an Olympus BX61W microscope; all-reflective optics are employed throughout the line to reduce spectral aberrations. The laser source was a tunable Ti:Sapphire laser system (Coherent Chameleon Ultra II) coupled to an APE Compact optical parametric oscillator, which provides a spectral range from $0.2$ to $3.7$ eV, enabling probing of bot near-conduction and near-valence band tail states. A homebuilt difference frequency generation line was employed for probing of near conduction band states reported in Section III.E to acieve laser energies below 0.2 eV. Select measurements were also verified using a Fianium supercontinuum white light laser source coupled to a laser line tunable filter (Photon Etc.), which yields high-throughput characterization over the $0.7$ to $3.1$ eV range. Both setups are designed to maintain Poynting vector illumination stability on the TFT during the spectral scan.

A 52X cassegrain reflective objective is used to focus the laser onto a diffraction-limited spot centered on the TFT. The a-IGZO TFT active channel is operated under a forward bias of $\mathrm{V_{ON} + 5}$ V; the TFT is electrically connected to the measurement interface via a homebuilt electrical probe setup consisting of RF source-drain electrical probes. At each illumination wavelength, the photoconduction (PC) signal is retrieved from  the noise using a current pre-amplifier and a lock-in amplifier (Zurich HFLI). To eliminate the contribution of dark current background and hysteretic drift, we modulate the illumination source using an optical chopper at a frequency of $585$ Hz and reference the lock-in amplifier to this frequency. Simultaneous scanning photoconduction and back reflection maps are taken at each illumination energy using a 4f confocal scanning geometry to ensure uniformity of the illumination spot over the energy measurement range and across successive measurements. 

Figure 1(c) shows scanning back reflection (\textit{left}) and scanning photoconduction maps (SPCM, all other panels) for a back channel etch a-IGZO TFT upon photoexcitation at $h\nu=$1.8 eV. The PC images are colorized via a heatmap corresponding to the relative amplitude of the integrated trap density that is optically excited.  Line cuts as a function of gate voltages ($\mathrm{V_G}$) ranging from depletion-mode to enhancement-mode operation conditions are shown in Fig. 1(c). As the TFT gate voltage is scanned from depletion-mode ($\mathrm{V_G < 0}$ V) to enhancement-mode ($\mathrm{V_G > 0}$ V) operation,  the photoconduction amplitude becomes order of magnitude larger and becomes uniformly distributed across the active channel length. In depletion-mode operation ($\mathrm{V_G = -4}$ V), the device PC response is localized to the immediate area of the drain electrode, indicative of a Schottky-like barrier at the interface.  The amplitude of the spatially-mapped UPBC response can be directly mapped to generate the sub-gap DoS by spectrally scanning the energy of the power-normalized excitation laser from 0.3 to 4 eV. 

The photoconduction signal at each given photon energy ($\mathrm{h \nu}$) is directly proportional to the total integrated subgap trap density. Specifically, the integrated trap density $\mathrm{N_{TOT}({h\nu})}$ between the conduction band mobility edge and $h\nu$ is given by:

\begin{equation}
\mathrm{N_{TOT}(h\nu)} =  \left(  \mathrm{\frac{\mathrm{qN_{o}}C_{I}}{t m}} \right)    \mathrm{\frac{I_{PC}}{N_{ph}}}
\end{equation} where $\mathrm{I_{PC}}$/$\mathrm{N_{ph}}$ represents the PC signal ($\mathrm{I_{PC}}$) which has been photon normalized via through-chip laser power correction at each photon energy, $\mathrm{h\nu}$. The bracketed quantity is a scaling constant composed of the the electronic charge (q), the gate insulator capacitance density($\mathrm{C_{I}}$), the slope (m) of the non-illuminated $\mathrm{I_{D} - V_{G}}$ transfer curve within $\mathrm{\pm 0.5}$ V of the constant $\mathrm{V_{ON} + 5}$ V forward operating gate bias (taken immediately after each PC measurement), and the accumulation layer thickness ($\mathrm{t}$). Finally, $\mathrm{N_{o}}$ is a constant calibration term obtained by finding the saturation photon flux, corresponding to the maximum photon flux that yields a detectable change in the in the PC-signal detected for illumination just below the VB mobility edge. After directly mapping the total integrated trap density as a function of photon energy from the raw observable PC signal, we take its derivative to obtain the experimental density of states (DoS): $\mathrm{DoS({h\nu})}=\frac{d\mathrm{N_{TOT}}}{d\mathrm{({h\nu}})}$. Throughout this work, we refer to experiment subgap DoS$\mathrm{(E-E_C)}$ which is equivalent to DoS$\mathrm{(-h\nu)}$.

%Lastly, the transient response photoconduction (PC) response of select a-IGZO TFTs was taken using a Tektronix TDS 3054B 500 MHz digital oscilloscope, triggered by an external optical chopper.  The transient response kinetics traces were obtained as function of the laser excitation energy by scanning the laser energy from 0.8 to 3.2 eV, and the device rise and fall rates are extracted by a least squares deconvolution kit of the kinetics to a biexponetial function. 

\section{Results and Analysis}
\subsection{a-IGZO DoS by UBPC vs. ultrabroadband optical methods}
%As the gate voltage is swept to enhancement-mode conditions ($\mathrm{V_G = 0, 7}$ V, \textit{bottom} panels), the PC becomes broadly distributed across the entire channel area.

   \begin{figure}
   \begin{center}
   \begin{tabular}{c}
   \includegraphics[height=12.5cm]{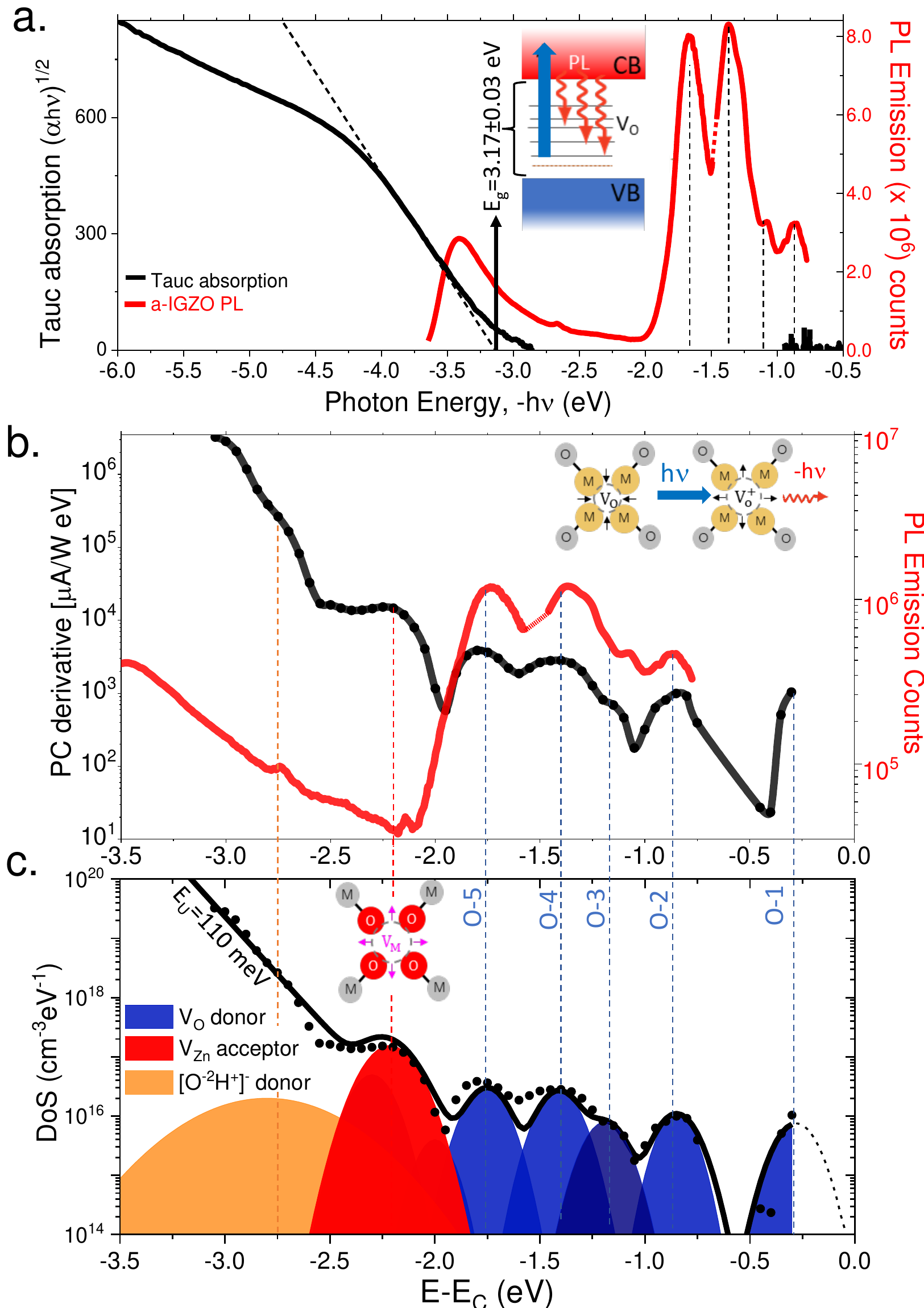}
   \end{tabular}
   \end{center}
   \caption[example] 
%>>>> use \label inside caption to get Fig. number with \ref{}
   { \label{fig:fig5pt2}  \textit{Ultrabroadband PL emission and photoconduction spectra, both correlated to a-IGZO subgap density of states peaks.}  (\textbf{a}) Tauc absorption (\textit{black}) and photoluminescence (PL, \textit{red}) emission spectra. (\textit{Inset}) PL emission is due to excited conduction band electrons recombining into deep trap states. (\textbf{b}) a-IGZO thin film PL emission spectrum (\textit{red}) is plotted on the same axis as the energy derivative of photoconduction (PC) spectrum of a back channel etch (BCE) a-IGZO TFT. Dotted lines highlight correspondence between PL and UBPC peaks. (\textbf{c}) UBPC density of states for a BCE a-IGZO TFT.}
   \end{figure} 
% The VB tail Urbach energy of is $\mathrm{110}$ meV for the BCE TFT and $\mathrm{414 \pm 2}$ meV for the thin film
%(\textit{left} scale, \textit{black}) (\textit{right} scale; \textit{dark red}

%(\textit{blue}: $\mathrm{V_O}$ donor; \textit{magenta}: $\mathrm{V_M}$ acceptor; \textit{orange}: $\mathrm{{[{O_{O}^{2-}}{H^+}]}^{1-}}$ donor).%
The predictive trends the UBPC DoS method is independently verified
by overlaying prominent DoS peaks with the a-IGZO photoluminescence emission spectrum collected over an ultrabroadband spectral range of 0.7 to 4 eV. Figure 2(a) plots Tauc\cite{Tauc1968} absorption $\mathrm{({\alpha \nu}^{1/2})}$ and photoluminescence (PL) emission spectra. The estimated optical bandgap ($\mathrm{E_g}$) for a-IGZO from the Tauc plot is $\mathrm{3.17 \pm 0.03}$ eV. PL reveals two strong subgap peaks at $\mathrm{\sim -1.4}$ and $\mathrm{\sim -1.7}$ eV, and two weak subgap peaks at $\mathrm{\sim -0.8}$ and $\mathrm{\sim -1.1}$ eV. The apparent drop-off in emission at $\mathrm{\sim -3.5}$ eV is actually an artifact due to PL drop off associated with use of a cut-off filter used to eliminate higher-harmonics in the monochromator.

The four subgap PL peaks (\textit{red}) of Fig. 2(a), as obtained from a 100 nm a-IGZO thin film, are plotted on an expanded energy scale in Fig. 2(b) and are shown to be correlated to photoconduction (PC) derivative ($\mathrm{\mu A W^{-1} eV^{-1}}$) spectrum peaks (\textit{black}), as obtained from a BCE a-IGZO TFT. As evident from Fig. 2(b), the match between PL and PC derivative peaks is quite striking. Three other PC derivative spectral features are indicated in Fig. 2(b) at $\mathrm{\sim -0.3}$, $\mathrm{\sim -2.2}$, and $\mathrm{\sim -2.8}$ eV.

Atomic identification of the PL and PC derivative peaks shown in Fig. 2(b) is shown in Fig. 2(c) and summarized in Table I. In Fig. 2(c), raw UBPC data (\textit{black} closed circles) is simulated (\textit{black} solid line) by convolving a series of Gaussian subgap defect peaks and an exponential valence band Urbach\cite{Urbach1953} tail. The UBPC spectrum indicated in Fig. 2(c) corresponds to the BCE a-IGZO TFT density of states parameters collected in Table I. UBPC density of states parameters for a top-gate (TG) a-IGZO TFT are also included in Table I, and are expected to be representative of the 100 nm a-IGZO thin film used for PL assessment, as both a-IGZO films are prepared in a similar manner.

 %Figure 2(b) (\textit{inset})show s graphical representation of the radiative relaxation of a $\mathrm{V_{O}}$ donor defect site by electron capture by an empty (positively ionized) charge site shown. Loss of a trapped electron from a $\mathrm{V_{O}}$ defect means that the positive charge of the (unoccupied) defect is no longer screened, causing nearby metal atoms to orient themselves away from the ionized defect site.

%The (\textit{inset}) of Fig. 2(a) depicts the mechanism for the PL emission measurement: the UV excitation source liberates trapped electrons across the entire subgap into the conduction band (CB), from which they recombine to lower-lying $\mathrm{V_O}$ or $\mathrm{V_{Zn}}$ states, emitting a photon whose energy corresponds to the energetic location of the defect recombination center in the subgap. The photon energy ($\mathrm{h \nu}$) notation is employed interchangeably with $\mathrm{E – E_{C}}$ (expressing an energetic location in the a-IGZO subgap) and is therefore represented on a negative axis.

%Figure 2(b) compares the energy derivative of the raw PC observable ($\mathrm{\mu A/W}$ eV) for a BCE a-IGZO TFT to the photoluminescence (PL) emission spectrum of a 100 nm a-IGZO thin film. 

%At $\mathrm{\sim -2.3}$ eV below the CB mobility edge, the subgap states often also contain acceptor-like Zn vacancy states shown by the red shading in Fig. 2(b-c).\cite{Vogt2020} 

Several aspects of Fig. 2(c) are notable. First, the four PL peaks (\textit{red}) of Fig. 2(b) are ascribed to oxygen vacancies, O-2, O-3, O-4, and O-5 (see Table I). Second, the strong peak occurring at -2.2 eV in both the PC derivative and UBPC spectra is attributed to a zinc vacancy acceptor, Zn-8. A zinc vacancy has previously been identified as the only cation vacancy energetically favored to exist in a-IGZO.\cite{Vogt2020} Also, recently reported electron spin resonance measurements confirm the existence of a zinc vacancy in a-IGZO at concentrations up to about an order of magnitude greater than that of estimated peak oxygen vacancy concentrations.\cite{Park2022} Note the absence of a -2.2 eV zinc vacancy peak in the PL spectrum of Fig. 2(b), consistent with the low concentration of this peak in Table I for a TG a-IGZO TFT. Third, the PC derivative and UBPC spectra show evidence of the O-1 oxygen vacancy peak at -0.3 eV. Fourth, the PC derivative curve shows evidence of an $\mathrm{{[{O_{O}^{2-}}{H^+}]}^{1-}}$ (or $\mathrm{OH^{-}}$) defect complex state at -2.8 eV, while this state is not observed in the UBPC spectrum, presumably because it is obscured by the valence band Urbach band tail. Simulation of the UBPC spectrum reveals a valence band Urbach energy, $\mathrm{E_U = 110}$ meV.

%The $\mathrm{V_{Zn}}$ states have generally attracted less interest as they shallow donor region where the enhancement-mode operation of AOS TFTs is typically mediated. 
% NOT TURE ANYORE: In succeeding sections, we will explore cases in which Zn vacancy states can substantially impact AOS TFT device operation.

\begin{figure*}
   \begin{center}
   \begin{tabular}{c}
   \includegraphics[height=13cm]{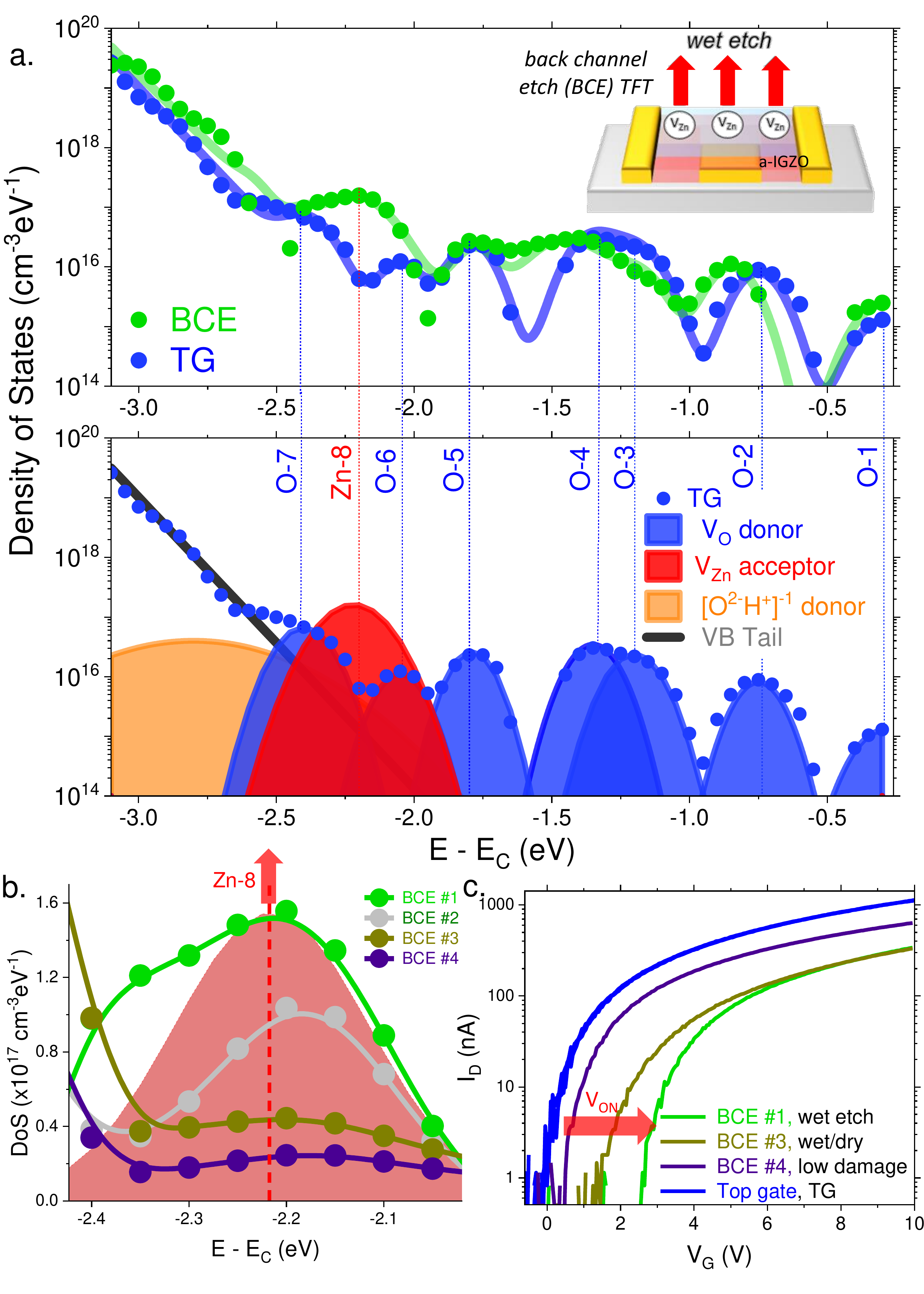}
   \end{tabular}
   \end{center}
   \caption[example] 
%>>>> use \label inside caption to get Fig. number with \ref{}
   { \label{fig:fig5pt3} \textit{Density of states (DoS) trends for a top-gate (TG) vs. a back channel etch (BCE) a-IGZO TFT.}
 \textbf{(a)} (\textit{Upper}) UBPC comparison of a TG and a BCE a-IGZO TFT. (\textit{Lower}) Simulated Gaussian subgap state peaks corresponding to the UBPC experimental data. \textbf{(b)} Linear DoS scaling highlights the Zn-8 zinc vacancy peak (\textit{red}) for different BCE processing conditions: \textit{1}: Mo S/D wet etch, \textit{2}: Ti/Cu S/D wet etch, \textit{3}: Ti/Cu S/D wet/dry etch, and \textit{4}: Ti/Cu S/D low damage etch. \textbf{(c)}  Drain current-gate voltage ($\mathrm{I_D - V_G}$) transfer curves for different BCE processing conditions.}
   \end{figure*}

\begin{center}
\begin{table*}
 \begin{tabular}{c | c  c|  c  c | c c} 
%\toprule[1.5pt]
Defect &  Peak Energy (eV)  & &  Peak DoS ($\mathrm{\times 10^{15}  cm^{-3}} \mathrm{eV^{-1}}$)& & FWHM (meV) \\
%, $\mathrm{E_{T}}$ $\mathrm{\omega}$
 \hline
 &  BCE & TG & BCE &  TG & BCE &  TG \\ [0.5ex] 

 \hline

O-1 & $\mathrm{-0.3}$ & $\mathrm{-0.3}$ & $\mathrm{2.5}$ & $\mathrm{1.3}$ & $\mathrm{80}$ & $\mathrm{80}$ \\

O-2 & $\mathrm{-0.85}$ & $\mathrm{-0.75}$ & $\mathrm{11}$ & $\mathrm{8.8}$ & $\mathrm{70}$ & $\mathrm{70}$ \\

O-3 & $\mathrm{-1.2}$ & $\mathrm{-1.2}$ & $\mathrm{24}$ & $\mathrm{9.2}$ & $\mathrm{80}$ & $\mathrm{80}$ \\

O-4 & $\mathrm{-1.35}$ & $\mathrm{-1.45}$ & $\mathrm{37}$ & $\mathrm{30}$ & $\mathrm{110}$ & $\mathrm{80}$ \\

O-5 & $\mathrm{-1.8}$ & $\mathrm{-1.8}$ & $\mathrm{23}$ & $\mathrm{27}$ & $\mathrm{70}$ & $\mathrm{70}$ \\

O-6 & $\mathrm{-2.0}$ & $\mathrm{-2.0}$ & $\mathrm{4}$  & $\mathrm{12}$ & $\mathrm{70}$ & $\mathrm{70}$ \\

O-7 & $\mathrm{-2.4}$ & $\mathrm{-2.4}$ & $\mathrm{50}$ & $\mathrm{67}$ & $\mathrm{50}$ & $\mathrm{80}$ \\

{Zn-8} & $\mathrm{-2.2}$ & $\mathrm{-2.2}$ & $\mathrm{\textbf{155}}$  & $\mathrm{\textbf{0.1}}$ & $\mathrm{130}$ & $\mathrm{130}$\\

%Zn-8 & $\mathrm{-2.6}$ & $\mathrm{-2.6}$ & $\mathrm{110}$ & $\mathrm{0.1}$ & $\mathrm{0.15}$\\
 %\hline
$\mathrm{{[{O_{O}^{2-}}{H^+}]}^{1-}}$ & $\mathrm{-2.8}$ & $\mathrm{-2.8}$ & $\mathrm{20}$ & $\mathrm{40}$ & $\mathrm{320}$ & $\mathrm{320}$ \\

%\bottomrule[1.5pt]

 \end{tabular}
  \caption[example]
 { \label{table1} 
Extracted parameters from UBPC measurements on the BCE and TG a-IGZO TFTs plotted in Fig. 3a show sub-gap defect peak energy ($\mathrm{E-E_{C}}$), peak density of states (DoS), and peak full-width-half-maximum (FWHM) are similar with the exception of the Zn-8 deep acceptor row (\textit{in bold}).
}
 \end{table*}
\end{center}

\subsection{Top-gate (TG) vs. Back Channel Etch (BCE) a-IGZO TFT Trends}
%Figure 3(a) (\textit{Panel} (1)) illustrates a common fabrication process for a back channel etch (BCE) a-IGZO TFT. A gate is first patterned onto a glass substrate prior to deposition of the gate oxide ($\mathrm{SiO_x}$), followed by the a-IGZO active channel layer; a metal layer is then deposited on the active layer to enable patterning of the source and drain electrodes. Fig. 3(a) (\textit{Panel} (2)) depicts the removal of this top material from the region over the active channel via a wet etch process.
%It has also been noted that hydrogen can be incorporated during efforts to passivate BCE a-IGZO TFTs.\cite{Nguyen2015}

Figure 3(a, \textit{upper}) compares UBPC trends (\textit{circles}: experimental data, \textit{solid curves}: simulation) for a TG a-IGZO TFT (\textit{blue}, \textit{solid} circles) to that of a BCE a-IGZO TFT (\textit{green}, \textit{solid} circles). Figure 3(a, \textit{lower}) plots the UBPC experimental data of a TG a-IGZO TFT ($\mathrm{solid}$ circles) with a simulation of the UBPC spectra based on Gaussian peaks convolved with an exponentially decaying valence band Urbach tail. Simulated subgap peaks are enumerated according to species ($\mathrm{V_O}$ donor, $\mathrm{V_{Zn}}$ acceptor, or $\mathrm{{[{O_{O}^{2-}}{H^+}]}^{1-}}$ donor) and by peak energetic location (from CB $\mathrm{\sim 0}$ eV to VB $\mathrm{\sim -3.2}$ eV). The oxygen vacancy (\textit{blue}) and $\mathrm{{[{O_{O}^{2-}}{H^+}]}^{1-}}$ (\textit{orange}) simulated peaks plotted in Fig. 3(a) correspond to simulation fitted to a TG UBPC spectrum, while the $\mathrm{V_{Zn}}$ peak (\textit{red}) corresponds to simulation fitted to a BCE UBPC spectrum. The characteristic energy of the VB tail exponential decay (or Urbach energy) of the UBPC spectrum is derived from the simulation of the experimental data by fixing all Gaussian subgap state peak energies and amplitudes, leaving the Urbach energy (TG: 89 meV, BCE: 95 meV) as the only free simulation parameter. The fitted Gaussian peak characteristics for the enumerated peaks are provided in Table I.  %($\mathrm{E_T}$ peak central energy (eV): peak maximum amplitude ($\mathrm{\times 10^{15}}$ $\mathrm{{cm^{-3}}{{eV}^{-1}}}$), $\mathrm{\omega}$: full width half-maximum (eV)) 

In Fig. 3(b), the UBPC DoS in the region of the Zn-8 zinc vacancy peak is compared on a linear ordinate scale for several different BCE processing conditions: \textit{1}: Ti/Cu S/D wet etch; \textit{2}: Mo S/D wet etch; \textit{3}: Ti/Cu S/D wet/dry etch (wet etch of the S/D metal far from the back channel surface, followed by dry etch near the back channel surface); and \textit{4}: Ti/Cu S/D low damage etch. Figure 3(c) shows the drain current-gate voltage ($\mathrm{I_{D}-V_{G}}$) transfer curves for three of the TFTs plotted in Fig. 3(b), as well as for a TG a-IGZO TFT.

As clearly evident from Figs. 3(a) and 3(b), the Zn-8 zinc vacancy concentration is strongly affected by the type of BCE etchant used. Also, a larger Zn-8 zinc vacancy concentration leads to a more positive shift in the turn-on voltage. From a charge balance perspective, an increase in the deep acceptor concentration, i.e., Zn-8 zinc vacancies, is balanced by an increase in the concentration of ionized deep donors, i.e., oxygen vacancies, e.g., O-1, O-2, O-3, etc. Mathematically, $\mathrm{\Delta N_{DA} \approx \Delta N_{DA}(E_F)}$, where the deep donor Fermi level dependence recognizes that the Fermi level is modulated deeper into the bandgap, away from the conduction band mobility edge, until a sufficient density of oxygen vacancy donors is ionized in order to achieve charge neutrality.

The Zn-8 zinc vacancy concentration can be estimated by integration of its UBPC peak area or alternatively from the shift in the turn-on voltage of an $\mathrm{I_D - V_G}$ transfer curve via $\mathrm{N_{DA, ID-VG} \approx (C_I \Delta V_{ON}/q)^{3/2}}$, where $\mathrm{C_I}$ is the gate insulator capacitance density (9.7 $\mathrm{nF/{cm^{-2}}}$), q is the electronic charge, and $\mathrm{\Delta V_{ON} = \Delta V_{ON,BCE} - \Delta V_{ON,TG}}$.

\begin{center}
\begin{table}
 \begin{tabular}{c c |c | c c c} 

BCE & Etchant  & $\mathrm{\Delta V_{ON}}$ (V) & $\mathrm{N_{DA, ID-VG}}$  & $\mathrm{N_{DA, UBPC}}$ ($\mathrm{cm^{-3}}$) \\ [0.5ex] 

 \hline

\#1: & wet  & 2.55  & $\mathrm{6.1 \times 10^{16}}$ & $\mathrm{6.2 \times 10^{16}}$\\ 

\#3: &  wet/dry  & 1.35  & $\mathrm{2.3 \times 10^{16}}$ & $\mathrm{2.4 \times 10^{16}}$\\ 

\#4: & low damage  & 0.50 & $\mathrm{5.3 \times 10^{15}}$ & $\mathrm{1.0 \times 10^{16}}$\\

 \end{tabular}
  \caption[example]
 { \label{table2} 
Comparison of changes in deep acceptor (DA) trap density of of three BCE a-IGZO TFTs with different etch process conditions (from Fig. 3b). TFT $\mathrm{I_D - V_G}$ transfer curve turn-on voltage shifts ($\mathrm{\Delta V_{ON}}$) shift right to predict the DA trap density ($\mathrm{N_{DA, ID-VG}}$) that is also measured directly by the UBPC method ($\mathrm{N_{DA, UBPC}}$).  
}
 \end{table}

\end{center}
Table II compares values of $\mathrm{N_{DA, ID-VG}}$ and  $\mathrm{N_{DA, UBPC}}$ for the three BCE a-IGZO TFTs plotted in Fig. 2(d). Note that the UBPC estimate is invariably larger, as UBPC measures non-electrically active valence band tail donor-like traps.\cite{Vogt2020,Mattson2022}

The UBPC spectra provided in Fig. 2(b-c) suggest that: (i). a standard BCE wet etch process results in the formation of a Gaussian zinc vacancy state centered at $\mathrm{\sim -2.2}$ eV (Zn-8), as well as a slight increase in the Urbach energy; (ii). zinc vacancy state formation effect is particularly severe for BCE TFTs with Mo source-drain electrodes; and (iii). the zinc vacancy concentration can be reduced through the use of less-damaging etch methods. The presence of this zinc vacancy peak and its particular susceptibility to form in a-IGZO TFTs processed using a Mo S/D wet etch is consistent with previously published work\cite{Park2007, Kwon2010, Nag2014, Park2015} on BCE etch damage. In Supplemental Materials Fig. S1, we provide the UBPC DoS for 6 BCE and 2 TG a-IGZO TFTs subjected to various processing conditions, along with the averaged BCE and TG DoS across all the devices studied. For all BCE devices, the Zn-8 peak is larger than that observed for any of the TG TFTs.

A number of strategies have been explored to mitigate deleterious effects of the etch process on BCE a-IGZO TFT electrical operation: careful selection of source/drain metal;\cite{Nag2014} plasma treatments intended to reduce metal residues;\cite{Liu2015} the use of etch stop layers (ESL);\cite{Kwon2010, Li2013, Nag2014, Luo2015} less damaging wet etchants such as $\mathrm{{H_2}{O_2}}$;\cite{Ryu2012} or through adopting more complex device architectures such as a dual-gate TFT.\cite{Nag2017} A comparison of an as-grown BCE TFT to a BCE TFT with an ESL is included in Supplemental Materials Fig. S2. UBPC measurements on ESL device show a smaller zinc vacancy DoS peak and improved electrical performance compared to an as-grown TFT.

\subsection{Plasma treatments of a-IGZO TFTs}

\begin{figure}
   \begin{center}
   \begin{tabular}{c}
   \includegraphics[height=14.5cm]{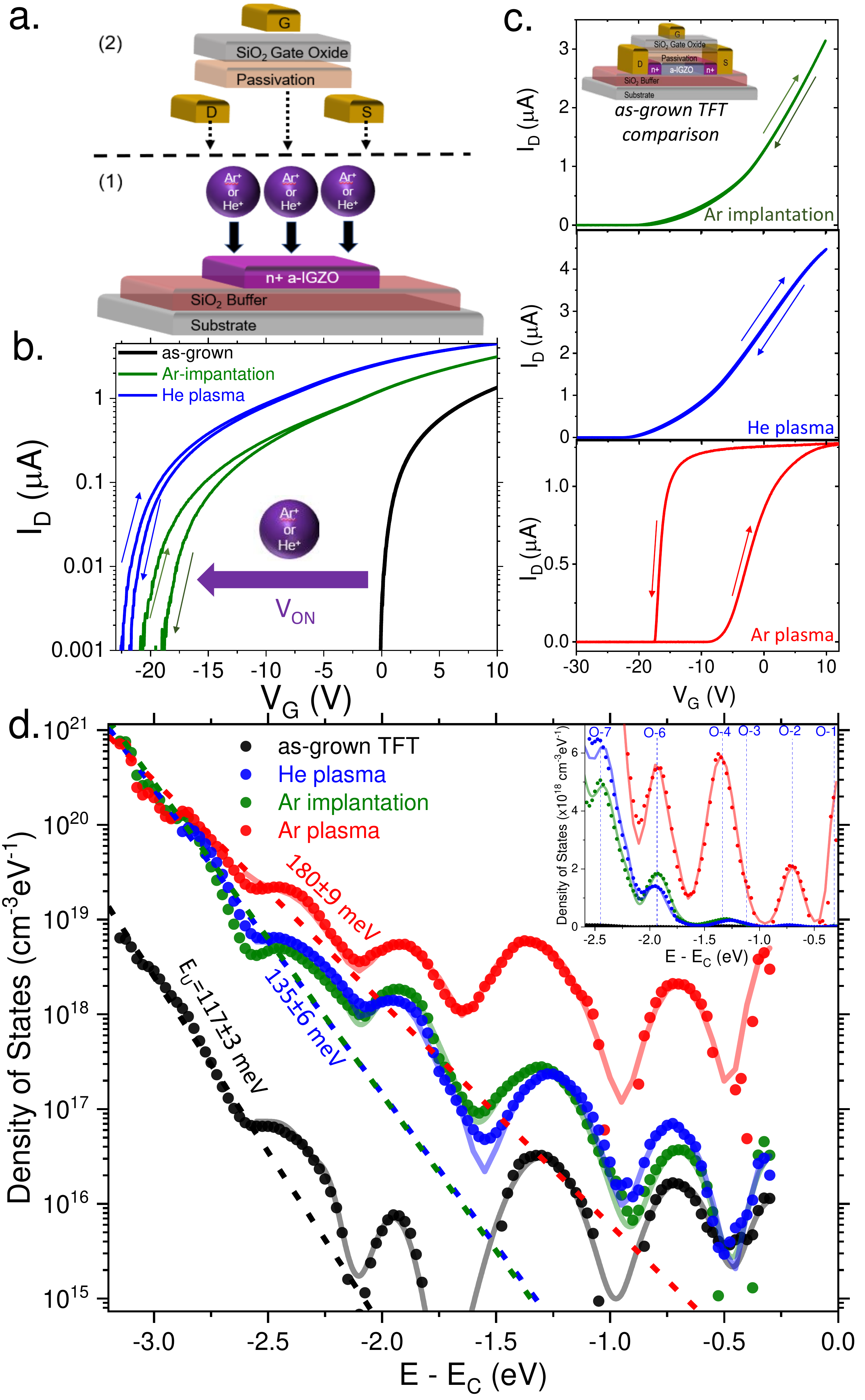}
   \end{tabular}
   \end{center}
   \caption[example] 
%>>>> use \label inside caption to get Fig. number with \ref{}
   { \label{fig:fig5pt4} 
\textit{Plasma processing of a-IGZO TFTs.} \textbf{(a)} $\mathrm{n^{+}}$ doping of the active channel of a top-gate a-IGZO TFT via plasma treatment or ion implantation prior to deposition of the top-gate electrode. 
%MOVE TO MAIN text: (\textit{Panel} (1)) An a-IGZO channel deposited on a glass substrate with an $\mathrm{SiO_2}$ buffer layer is subjected to a plasma or ion implantation treatment, resulting in an n+ a-IGZO active channel. (\textit{Panel} (2)) Source-drain electrodes, and top-gate electrode are patterened or deposited onto the n+ a-IGZO active channel to form the n+ a-IGZO active channel TFT. 
\textbf{(b)} The drain current-gate voltage ($\mathrm{I_{D}-V_{G}}$) transfer curve of an as-grown top-gate a-IGZO TFT compared to those subjected to an $\mathrm{Ar^+}$ ion implantation or He plasma active channel pre-treatment. \textbf{(c)} The $\mathrm{I_{D}-V_{G}}$ transfer curves of three $\mathrm{n^{+}}$ a-IGZO active channel TFTs subjected to different plasma treatments. \textbf{(d)} Subgap density of states ($\mathrm{cm^{-3}eV^{-1}}$) of as-grown and plasma-treated top-gate TFTs whose $\mathrm{I_{D}-V_{G}}$ transfer curves) are plotted in (b-c). (\textit{Inset}) Linear ordinate scaling of DoS in the oxygen vacancy  region of the subgap.}
   \end{figure} 

Figure 4(a) depicts the fabrication of a $\mathrm{n^{+}}$ active channel TG a-IGZO TFT via plasma or ion implantation treatment. In (\textit{Panel} (1)), an a-IGZO channel deposited onto a glass substrate with a $\mathrm{SiO_2}$ buffer layer is subjected to a plasma or ion implantation treatment, resulting in a $\mathrm{n^{+}}$ a-IGZO active channel. In (\textit{Panel} (2)), source-drain electrodes and the top-gate are patterned or deposited onto the $\mathrm{n^{+}}$ a-IGZO active channel to form the $\mathrm{n^{+}}$ a-IGZO active channel TFT. Note that in this process, the entire active channel is subjected to the $\mathrm{n^{+}}$ doping treatment.

In Fig. 4(b), $\mathrm{log(I_{D})-V_{G}}$ transfer curves are compared for three TG a-IGZO TFTs in which the channel is as-grown (\textit{black}) or is subjected to an $\mathrm{Ar^+}$ ion implantation (\textit{green}) or a He plasma (\textit{blue}) treatment to obtain a $\mathrm{n^{+}}$ a-IGZO active channel. \textit{Arrows} represent the direction of hysteresis. Compared to the as-grown a-IGZO TFT, the plasma-processed a-IGZO TFTs are strongly depletion-mode, with large negative turn-on voltages of $\mathrm{< -20}$ V. 

%(Here, we define $\mathrm{V_{ON}}$ the voltage at which the drain current rises above the measurement noise floor for the forward-going gate voltage sweep.)
 
Figure 4(c) plots the $\mathrm{I_{D}-V_{G}}$ transfer curves of three TG a-IGZO TFTs whose active channel is subjected to treatment by $\mathrm{Ar^+}$ ion implantation (\textit{green}), He plasma (\textit{blue}), or Ar plasma (\textit{red}). The $\mathrm{Ar^+}$ ion-implanted and He plasma-treated TFTs exhibit no measurable hysteresis when plotted on a linear ordinate scale (Fig. 4(c)), and only a very small amount of clockwise hysteresis when plotted on a logarithmic scale (Fig. 4(b)). In contrast, the Ar plasma-treated TFT possesses a large amount of hysteresis and, notably, this hysteresis is counterclockwise. Typically, clockwise hysteresis in an n-channel TFT is ascribed to electron trapping, while counterclockwise hysteresis is attributed to ion migration.\cite{Wager2010} Thus, the counterclockwise hysteresis witnessed for the Ar plasma-treated TFT is tentatively attributed to ion migration. However, we note that this case is unusual since ion migration normally occurs within the gate insulator rather than the semiconductor.

The deleterious effects of plasma damage to the channel layer of an a-IGZO TFT are unambiguously revealed by UBPC, as shown in Fig. 4(d) and summarized in Table III. Plasma processing dramatically increases the amplitude of all six oxygen vacancy peaks. $\mathrm{Ar^+}$ ion implantation and He plasma treatments increase the amplitudes modestly of the shallow O-1 and O-2 deep traps by a factor of less than an order of magnitude. Ar plasma damage is the most dramatic, increasing the amplitudes of all of the oxygen vacancy subgap peaks by well over two orders of magnitude.

In addition to increasing the amplitude of the oxygen vacancy subgap peaks, the valence band Urbach energy increases after plasma processing. This Urbach energy increase is relatively modest for the $\mathrm{Ar^+}$ ion implantation or He plasma treatment - from 117 $\pm$ 3 meV (as-grown) to 133 $\pm$ 8  meV ($\mathrm{Ar^+}$ ion implantation) - and is more significant for the Ar plasma treatment at $\mathrm{E_U=}$ 180 $\pm$ 9 meV. The linear ordinate density of states plot included as an inset to Fig. 4(d) shows how much more damaging the Ar plasma treatment is than either the $\mathrm{Ar^+}$ ion implantation or the He plasma treatment. After Ar plasma treatment, the anion sublattice appears to be so heavily damaged that it is not surprising that a TFT channel layer subjected to such abuse would exhibit $\mathrm{I_{D}-V_{G}}$ transfer curve counterclockwise hysteresis due to ion migration (perhaps by hydrogen) on its sub-lattice.

Table III reports the strongest oxygen deep acceptor DoS peak ( $\mathrm{ \times 10^{16}}$ $\mathrm{cm^{-3}eV^{-1}}$) as obtained from the UBPC measurement data plotted in Fig. 3(d). As the $\mathrm{I_{D}-V_{G}}$ curves plotted in Fig. 3(c) indicate, all of the plasma-treated or ion-implanted TFTs experienced large negative shifts in the $\mathrm{I_{D}-V_{G}}$ curve $\mathrm{V_{ON}}$ curve ranging from $\mathrm{\sim -10-(-30)}$ V. This trend is consistent with $\mathrm{n^{+}}$ doping of the active channel via the $\mathrm{Ar^{+}}$ ion implantation or He/Ar plasma treatments.

A primary application of ion implantation or plasma treatment is the formation of $\mathrm{n^{+}}$ source/drain regions at the active channel edge; after plasma treatment (or ion implantation), these $\mathrm{n^{+}}$ doped regions function as low contact resistance source-drain regions. The UBPC DoS spectra and $\mathrm{I_{D}-V_{G}}$ transfer curves in Fig. 4 confirm that treatments that induce a moderate amount of disorder in the active channel material (forming shallow donor oxygen vacancy states) such as $\mathrm{Ar^{+}}$ ion implantation and He plasma are useful for achieving the desired $\mathrm{n^{+}}$ doping effect.

\begin{center}
\begin{table}
 \begin{tabular}{c  c |  c |c|  c|  c} 

 & $\mathrm{E-E_C}$(eV) & As-grown & $\mathrm{Ar^+}$ ion & He plasma & Ar plasma   \\ [0.5ex] 

 \hline

$\mathrm{O-1}$ & $-0.30$ & $1.1$ & $3.2 $ & $3.2 $ & $500$ \\ 

$\mathrm{O-2}$ & $-0.70$ &$1.7 $ & $3.7 $ & $7.0 $ & $210$  \\ 

$\mathrm{O-3}$ & $-1.30$ &$3.2 $ & $27 $ & $23 $ & $590 $  \\ 

$\mathrm{O-6}$ & $-1.95$ &$0.8 $ & $190 $ & $140 $ & $520 $ \\ 

$\mathrm{O-7}$ & $-2.45$ &$5.9$ & $420 $ & $500 $ & $1000 $ \\

 $\mathrm{E_U}$(meV) & &  $117 \pm 3$ & $133\pm8 $ & $135\pm6 $ & $180\pm9$ \\

 \end{tabular}
  \caption[example]
 { \label{table3} 
Comparison of density of states peak maximum amplitudes ($\times 10^{16} \mathrm{cm^{-3} eV^{-1 }}$) and Urbach energies (meV) from Fig. 4(d) for four top-gate a-IGZO TFTs with the full-channel subjected to different plasma and ion-etch processes.
}
 \end{table}

\end{center}

\begin{figure}
   \begin{center}
   \begin{tabular}{c}
   \includegraphics[height=7cm]{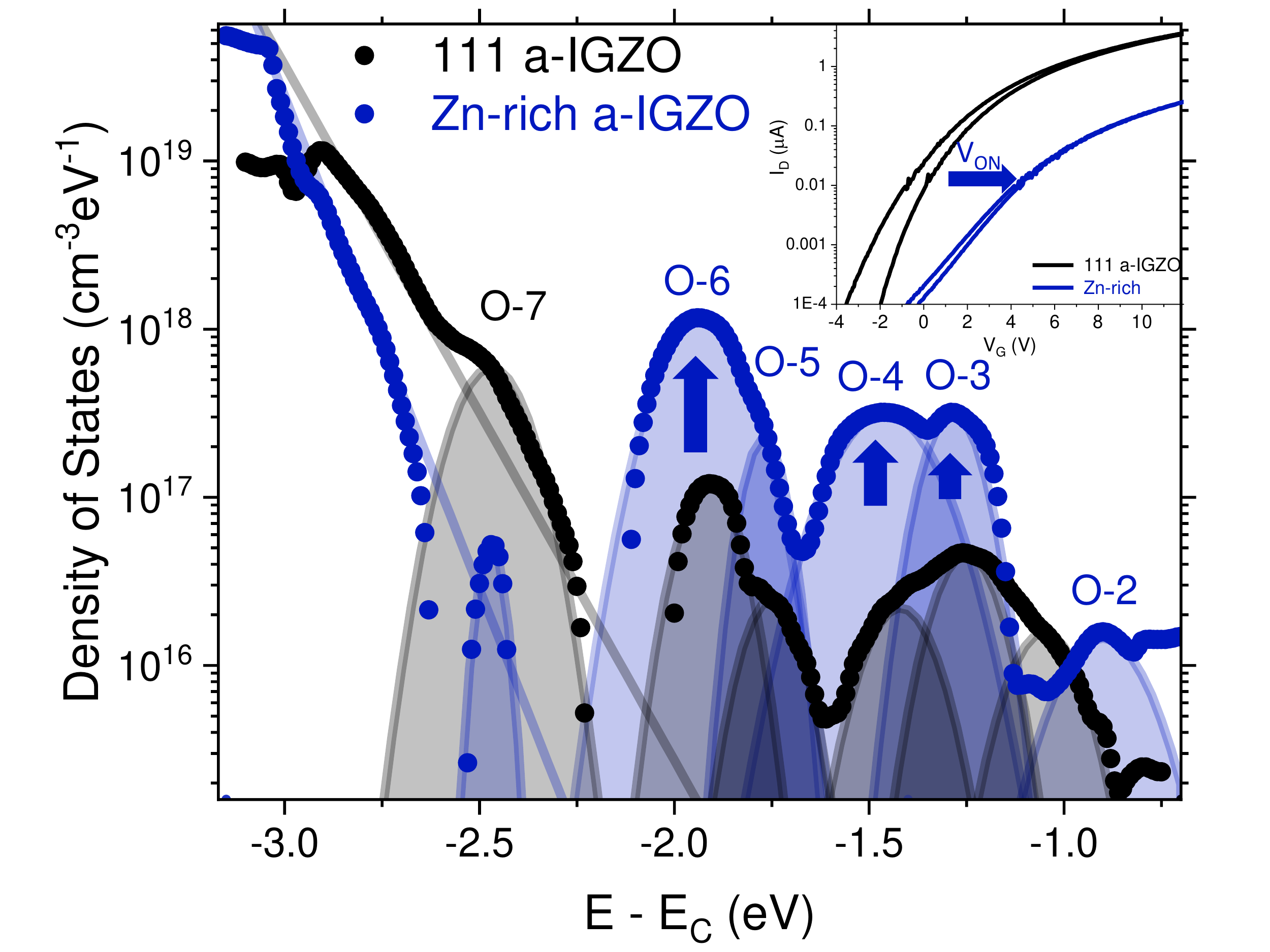}
   \end{tabular}
   \end{center}
   \caption[example] 
%>>>> use \label inside caption to get Fig. number with \ref{}
   { \label{fig:fig5pt5} UBPC DoS spectra comparing different a-IGZO growth conditions plotted for a BCE 111 (\textit{black}) and a BCE Zn-rich  (\textit{blue}) a-IGZO growths.  Fitted DoS peak are plotted as \textit{shaded} regions. (\textit{Inset}) Drain current-gate voltage ($\mathrm{I_{D}-V_{G}}$) transfer curves for a BCE 111 a-IGZO and a BCE Zn-rich a-IGZO TFT.}
   \end{figure} 

\subsection{a-IGZO channel stoichiometry variations}

\begin{center}
\begin{table}
 \begin{tabular}{c | c |  c |c } 

 & $\mathrm{E-E_C}$(eV) & 111 & Zn-rich  \\ [0.5ex] 

 \hline

$\mathrm{O-2}$ & $-1.05$ &$0.25 $ & $1.45 $  \\ 

$\mathrm{O-3}$ & $-1.25$ &$4.6 $ & $32 $  \\ 

$\mathrm{O-4}$ & $-1.40$ &$2.2 $ & $32 $  \\ 

$\mathrm{O-5}$ & $-1.75$ &$2.45 $ & $ 22 $  \\ 

$\mathrm{O-6}$ & $-1.90$ &$11.9 $ & $117 $  \\ 

$\mathrm{O-7}$ & $-2.50$ &$59$ & $5.2 $ \\

$\mathrm{E_U}$(meV) & &  $105$ & $78$ \\ 

 \end{tabular}
  \caption[example]
 { \label{table4} 
Comparison of the DoS peak maximum amplitudes ($\times 10^{16} \mathrm{cm^{-3} eV^{-1 }}$)  from Fig. 5(d) for [111] and Zn-rich growth recipes for BCE a-IGZO TFTs.
}
 \end{table}

\end{center}
 
Figure 5 plots the density of states (\textit{dots}: UBPC experimental data, \textit{shaded regions}: simulation) of a BCE a-IGZO TFT with a stoichiometric (1:1:1 molar \% $\mathrm{In_{2}O_{3}/Ga_{2}O_{3}/ZnO}$ ratio) active channel composition (111 a-IGZO, \textit{black}) against a BCE a-IGZO TFT with a higher molar proportion of Zn constituent atoms than In and Ga (Zn-rich, \textit{blue}). The (\textit{Inset}) of Fig. 5 plots the drain current-gate voltage ($\mathrm{I_{D}-V_{G}}$) transfer curve of the 111 as-grown and Zn-rich TFTs. Table IV reports the DoS peak characteristics for 111 and Zn-rich a-IGZO TFTs.

The DoS peak amplitudes of the O-3, O-4, O-5, and O-6 peaks are significantly larger for the Zn-rich a-IGZO TFT compared to the 111 a-IGZO TFT, but the O-7 peak is smaller. The $\mathrm{I_{D}-V_{G}}$ transfer curve of the Zn-rich a-IGZO TFT also exhibits a lower drain current, a positive shift in the turn-on voltage, and a poorer sub-threshold slope than the 111 a-IGZO TFT. These trends are consistent with an electron trapping-induced reduction of TFT electrical performance due to the enhanced density of subgap oxygen vacancy states. The appreciable increase in shallow oxygen vacancy states is particularly concerning with respect to TFT operation. Surprisingly, the valence band Urbach energy of the Zn-rich TFT is $\mathrm{\sim 30 \%}$ smaller than that of the 111 a-IGZO TFT.

\subsection{Hydrogen incorporation into a-ITGZO TFTs}
\begin{figure}
   \begin{center}
   \begin{tabular}{c}
   \includegraphics[height=9cm]{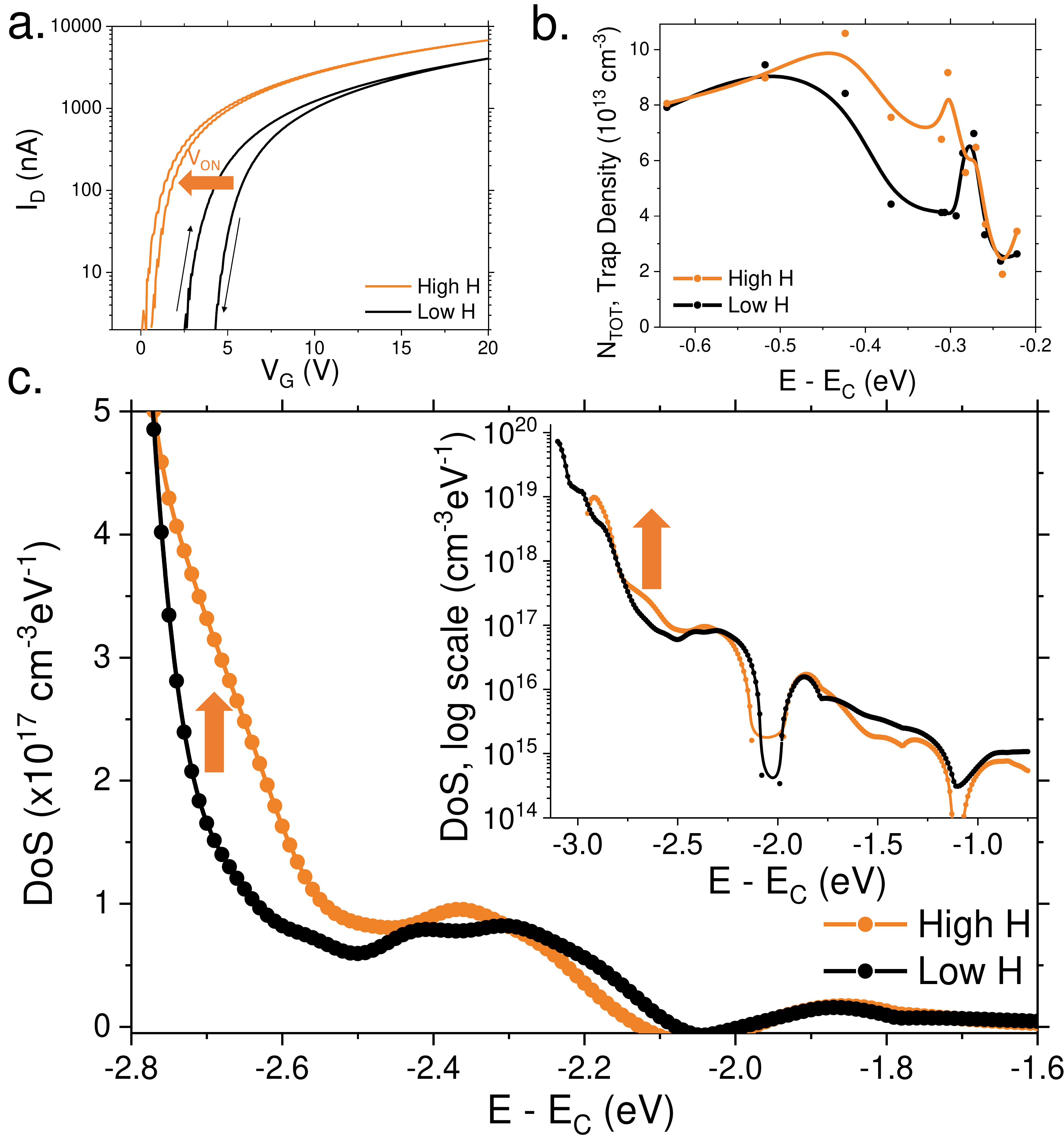}
   \end{tabular}
   \end{center}
   \caption[example] 
%>>>> use \label inside caption to get Fig. number with \ref{}
   { \label{fig:fig5pt6} 
\textbf{(a)} Drain current-gate voltage ($\mathrm{I_{D}-V_{G}}$) transfer curves, \textbf{(b)} near conduction band total integrated trap density, $\mathrm{N_{TOT}}$, and \textbf{(c)} near valence band DoS for two a-ITGZO BCE TFTs possessing different concentrations of hydrogen. (c, \textit{Inset}) UBPC DoS plotted across full bandgap (logarithmic scale).}
   \end{figure} 

Figure 6 displays a comparison between two a-ITGZO TFTs possessing different concentrations of hydrogen. $\mathrm{I_{D}-V_{G}}$ transfer curves (Fig. 6(a)) shows that the larger incorporated hydrogen concentration (\textit{orange} curve) shifts the turn-on voltage $\mathrm{-3.6}$ V, corresponding to an estimated increase in the hydrogen concentration of $\mathrm{\Delta[H]_{ID-VG} = 2.8 \times 10^{17}}$ $\mathrm{cm^{-3}}$, where $\mathrm{\Delta[H]_{ID-VG} = (C_{I}\Delta V_{ON} / q)^{1.5}}$ and $\mathrm{C_{I} = 19.2}$ $\mathrm{nFcm^{-2}}$ is the insulator capacitance density. (Since hysteresis is present in the $\mathrm{I_{D}-V_{G}}$ transfer curves shown in Fig. 6(a), $\mathrm{V_{ON}}$ is estimated as the midpoint between the right-going and left-going turn-on voltages.) The negative turn-on voltage shift indicates that the incorporated specie (hydrogen) behaves as a donor.

Since donor ionization energies are normally found nearest to the conduction band, Fig. 6(b) is included in order to demonstrate that the UBPC-estimated trap density is extremely small (less than $\mathrm{10^{14}}$ $\mathrm{cm^{-3}}$), and that the near conduction band trap distribution is quite similar for the two a-ITGZO BCE TFTs under consideration. These observations support our contention\cite{Mattson2022} that hydrogen does not behave as a normal shallow donor when incorporated into an amorphous oxide semiconductor, such as a-ITGZO.

Figure 6(c) reveals an enhanced density of states at $\mathrm{E - E_{C} \approx -2.7}$ eV for the high hydrogen (\textit{orange}) curve, compared to the low H (\textit{black}) curve. This corresponds to an increase in the concentration of $\mathrm{{[{O_{O}^{2-}}{H^+}]}^{1-}}$ defect complexes (or, equivalently, $\mathrm{OH^{-}}$) with increasing hydrogen incorporation. This increase in hydrogen between the \textit{black} and \textit{orange} curve can be quantified via UBPC, i.e., $\mathrm{\Delta [H]_{UBPC} = N_{TOT}(E_g)|_{\substack{\text{orange} \\ \text{curve}}} - N_{TOT}(E_g)|_{\substack{\text{black} \\ \text{curve}}}}$ $\mathrm{= 9.2 \times 10^{17} cm^{-3}}$ $\mathrm{- 6.0 \times 10^{17} cm^{-3}}$ $\mathrm{= 3.2 \times 10^{17} cm^{-3}}$. This value of $\mathrm{\Delta [H]_{UBPC}}$ ($\mathrm{3.2 \times 10^{17} cm^{-3}}$) is quite similar to that estimated previously for  $\mathrm{\Delta[H]_{ID-VG}}$ ($\mathrm{2.8 \times 10^{17} cm^{-3}}$), indicating that a correlation exists between the enhancement in density of states and the enhancement in electrically-active donor activity with increasing hydrogen concentration. However, note that $\mathrm{\Delta [H]_{UBPC} > \Delta[H]_{ID-VG}}$ since the UBPC-estimated density of states enhancement includes a contribution associated with valence band tail state density arising from enhanced anion sublattice disorder, as well as from a simple increase in the $\mathrm{{[{O_{O}^{2-}}{H^+}]}^{1-}}$ defect complex density centered at $\mathrm{E-E_{C} = -2.7}$ eV, due to direct incorporation of hydrogen into the amorphous network.\cite{Mattson2022}

Simple charge balance considerations are useful for rationalizing turn-on voltage trends similar to those presented in Fig. 6(a).\cite{Vogt2020} Enhancement-mode operation, in which $\mathrm{V_{ON}}$ is positive, is witnessed for the two curves included in Fig. 6(a). Charge balance describing this type of behavior likely arises from $\mathrm{N_{DD}^{+}(E_{F}) = N_{DA}-N_{SD}}$ where $\mathrm{N_{DD}^{+}}$ refers to positively ionized deep donors, the concentration of which is controlled by the position of the Fermi level, $\mathrm{N_{DA}}$ is the density of deep acceptors, and $\mathrm{N_{SD}}$ is the density of shallow donors. Identifying deep donors as oxygen vacancies that are distributed across the upper portion of the a-ITGZO bandgap (see \textit{Inset} to Fig. 6(c)), deep acceptors as zinc vacancies located at $\mathrm{E - E_{C} \approx -2.3-(-2.4)}$ eV, and shallow donors as $\mathrm{{[{O_{O}^{2-}}{H^+}]}^{1-}}$ defect complexes centered at $\mathrm{E - E_{C} \approx -2.8}$ eV, enhancement-mode behavior is a balancing act between zinc vacancies and hydrogen. Although $\mathrm{{[{O_{O}^{2-}}{H^+}]}^{1-}}$ defect complexes are certainly not energetically `shallow' since they are centered at $\mathrm{E - E_{C} \approx -2.8}$ eV, they can be considered to be `shallow' from the perspective of charge balance assessment since they remain ionized, independent of the position of the Fermi level. This odd behavior is due to the non-equilibrium nature of the hydrogen donor, in which hydrogen ionization occurs prior to its incorporation into the amorphous network.\cite{Mattson2022}

When the zinc vacancy density is much larger than that of incorporated hydrogen, $\mathrm{N_{DD}^{+}(E_{F}) \approx N_{DA}}$, so that strongly enhancement-mode behavior obtains (\textit{black} curve of Fig. 6(a)), $\mathrm{V_{ON}}$ is positive and large, and $\mathrm{E_F}$ is positioned deep in the gap (perhaps $\mathrm{-1.5}$ to $\mathrm{-2}$ eV), leaving many oxygen vacancy traps empty (empty traps ($\mathrm{N_{DD}^{+}}$) are responsible for enhancement-mode behavior). In contrast, when the zinc vacancy and incorporated hydrogen concentrations are similar, $\mathrm{N_{D}^{+}(E_{F}) \approx 0}$, such that $\mathrm{V_{ON} \approx 0}$ V, and $\mathrm{E_F}$ is positioned about $\mathrm{0.15-0.3}$ eV below the conduction band mobility edge, and most oxygen vacancy traps are filled. Finally, if the incorporated hydrogen concentration exceeds that of the zinc vacancy concentration, then depletion-mode behavior occurs, as described by a different charge balance relationship, $\mathrm{n(E_{F}) \approx N_{SD} - N_{DA}}$, where n is the free electron density, as controlled by the position of the Fermi level.

\section{CONCLUSIONS}
The subgap DoS measured by ultrabroadband photoconduction (UBPC) method reveals systematic trends in AOS subgap trap density that correlate with different TFT architectures, doping treatments, and AOS growth compositions. Notably, we observed the following: \textit{1}, a Zn vacancy peak centered at $\mathrm{-2.2}$ eV is introduced to back channel etch a-IGZO TFTs as a result of the source/drain metal etch process, and its peak density can be modulated via different etch conditions; \textit{2}, Ar and He plasma and $\mathrm{Ar^{+}}$ ion implantation-treatment of the a-IGZO active channel enhance the peak densities of deep-donor oxygen vacancy states up to $\mathrm{>100}$X; \textit{3}, Zn-rich a-IGZO TFTs exhibit larger deep oxygen vacancy peaks and significantly worse electrical performance compared to a-IGZO with a 1:1:1 stoichiometric molar proportion of constituent cations; and \textit{4},  as previously observed for a-IGZO, \cite{Mattson2022} hydrogen incorporation in a-ITGZO rigidly shifts the $\mathrm{I_{D}-V_{G}}$ transfer curve negative, and is accompanied by a corresponding increase in the $\mathrm{{[{O_{O}^{2-}}{H^+}]}^{1-}}$ complex peak density centered at $\mathrm{-2.8}$ eV.

 The defect DoS obtained with the UBPC method has both high 10$^6$ signal to noise sensitivity and a broad energy range not yet obtainable by  established purely optical and XPS/UPS methods.  This newly demonstrated capacity of UBPC to predict defects arising from on-chip TFT processing and growth recipes further suggests its use as an analytical tool in AOS TFT development. The different categories of defects (acceptor, donor, and interstitial hydrogen) that compose the subgap of AOS impact both the TFT $\mathrm{I_{D}-V_{G}}$ transfer curve turn-on voltage and hysteresis. The next challenge is to extend the UBPC method over the conduction band Urbach tails to complete the connection between subgap trap density and performance TFT metrics. The clear DoS trends with TFT production processing demonstrate that UBPC is an emerging post-assembly defect characterization method that can work down to the single-pixel limit.

\begin{acknowledgments}
\textbf{Acknowledgments}:  This work was in part supported by the NSF Grants DMR-1920368.  We would  like to thank Jessica Waymire for her contribution to the optical photoluminescence of a-IGZO measurements. 
\end{acknowledgments}

\textbf{Supporting Information Available}: Supplemental experiment DoS curves showing a statistical comparison between many TFT amorphous oxide devices.

\bibliography{bibtex}   %>>>> bibliography data in report.bib

\begin{thebibliography}{10}
\providecommand{\url}[1]{\texttt{#1}}
\providecommand{\urlprefix}{URL }

\bibitem{Kamiya2010}
T.~Kamiya, K.~Nomura, H.~Hosono,
\newblock \emph{Science and Technology of Advanced Materials} \textbf{2010},
  \emph{11}, 4 044305.

\bibitem{Wager2014}
J.~F. Wager, B.~Yeh, R.~L. Hoffman, D.~A. Keszler,
\newblock \emph{Current Opinion in Solid State and Materials Science}
  \textbf{2014}, \emph{18}, 2 53.

\bibitem{Nomura2004}
K.~Nomura, H.~Ohta, A.~Takagi, T.~Kamiya, M.~Hirano, H.~Hosono,
\newblock \emph{Nature} \textbf{2004}, \emph{432}, 7016 488.

\bibitem{Cho2011}
E.~N. Cho, J.~H. Kang, C.~E. Kim, P.~Moon, I.~Yun,
\newblock \emph{{IEEE} Transactions on Device and Materials Reliability}
  \textbf{2011}, \emph{11}, 1 112.

\bibitem{Ahn2014}
B.~D. Ahn, H.-S. Kim, D.-J. Yun, J.-S. Park, H.~J. Kim,
\newblock \emph{{ECS} Journal of Solid State Science and Technology}
  \textbf{2014}, \emph{3}, 5 Q95.

\bibitem{Vogt2020}
K.~T. Vogt, C.~E. Malmberg, J.~C. Buchanan, G.~W. Mattson, G.~M. Brandt, D.~B.
  Fast, P.~H.-Y. Cheong, J.~F. Wager, M.~W. Graham,
\newblock \emph{Physical Review Research} \textbf{2020}, \emph{2}, 3.

\bibitem{Kim2015}
K.-A. Kim, M.-J. Park, W.-H. Lee, S.-M. Yoon,
\newblock \emph{Journal of Applied Physics} \textbf{2015}, \emph{118}, 23
  234504.

\bibitem{Yoon2018}
S.-J. Yoon, N.-J. Seong, K.~Choi, W.-C. Shin, S.-M. Yoon,
\newblock \emph{{RSC} Advances} \textbf{2018}, \emph{8}, 44 25014.

\bibitem{Zhang2019}
Y.~Zhang, H.~Xie, C.~Dong,
\newblock \emph{Micromachines} \textbf{2019}, \emph{10}, 11 779.

\bibitem{Fung2009}
T.-C. Fung, C.-S. Chuang, C.~Chen, K.~Abe, R.~Cottle, M.~Townsend, H.~Kumomi,
  J.~Kanicki,
\newblock \emph{Journal of Applied Physics} \textbf{2009}, \emph{106}, 8
  084511.

\bibitem{Chen2011}
W.-T. Chen, S.-Y. Lo, S.-C. Kao, H.-W. Zan, C.-C. Tsai, J.-H. Lin, C.-H. Fang,
  C.-C. Lee,
\newblock \emph{{IEEE} Electron Device Letters} \textbf{2011}, \emph{32}, 11
  1552.

\bibitem{Song2016}
J.~H. Song, N.~Oh, B.~D. Anh, H.~D. Kim, J.~K. Jeong,
\newblock \emph{{IEEE} Transactions on Electron Devices} \textbf{2016},
  \emph{63}, 3 1054.

\bibitem{Song2018}
H.~Song, G.~Kang, Y.~Kang, S.~Han,
\newblock \emph{physica status solidi (b)} \textbf{2018}, \emph{256}, 3
  1800486.

\bibitem{Mattson2022}
G.~W. Mattson, K.~T. Vogt, J.~F. Wager, M.~W. Graham,
\newblock \emph{Journal of Applied Physics} \textbf{2022}, \emph{131}, 10
  105701.

\bibitem{Urbach1953}
F.~Urbach,
\newblock \emph{Physical Review} \textbf{1953}, \emph{92}, 5 1324.

\bibitem{Wager2017}
J.~F. Wager,
\newblock \emph{AIP Advances} \textbf{2017}, \emph{7}, 12 1.

\bibitem{deJamblinnedeMeux2017}
A.~de~Jamblinne~de Meux, A.~Bhoolokam, G.~Pourtois, J.~Genoe, P.~Heremans,
\newblock \emph{physica status solidi (a)} \textbf{2017}, \emph{214}, 6
  1600889.

\bibitem{Wang2022}
C.~Wang, W.~Lu, F.~Li, H.~Ning, F.~Ma,
\newblock \emph{Journal of Applied Physics} \textbf{2022}, \emph{131}, 12
  125709.

\bibitem{Rajachidambaram2012}
J.~S. Rajachidambaram, S.~Sanghavi, P.~Nachimuthu, V.~Shutthanandan, T.~Varga,
  B.~Flynn, S.~Thevuthasan, G.~S. Herman,
\newblock \emph{Journal of Materials Research} \textbf{2012}, \emph{27}, 17
  2309.

\bibitem{Du2014}
X.~Du, B.~T. Flynn, J.~R. Motley, W.~F. Stickle, H.~Bluhm, G.~S. Herman,
\newblock \emph{ECS Journal of Solid State Science and Technology}
  \textbf{2014}, \emph{3}, 9 Q3045.

\bibitem{Kim2007}
M.~Kim, J.~H. Jeong, H.~J. Lee, T.~K. Ahn, H.~S. Shin, J.-S. Park, J.~K. Jeong,
  Y.-G. Mo, H.~D. Kim,
\newblock \emph{Applied Physics Letters} \textbf{2007}, \emph{90}, 21 212114.

\bibitem{Kwon2010}
J.-Y. Kwon, K.~S. Son, J.~S. Jung, K.-H. Lee, J.~S. Park, T.~S. Kim, K.~H. Ji,
  R.~Choi, J.~K. Jeong, B.~Koo, S.~Lee,
\newblock \emph{Electrochemical and Solid-State Letters} \textbf{2010},
  \emph{13}, 6 H213.

\bibitem{Nag2014}
M.~Nag, A.~Bhoolokam, S.~Steudel, A.~Chasin, K.~Myny, J.~Maas, G.~Groeseneken,
  P.~Heremans,
\newblock \emph{Japanese Journal of Applied Physics} \textbf{2014}, \emph{53},
  11 111401.

\bibitem{Park2015}
Y.~C. Park, J.~G. Um, M.~Mativenga, J.~Jang,
\newblock \emph{{ECS} Journal of Solid State Science and Technology}
  \textbf{2015}, \emph{4}, 12 Q124.

\bibitem{Park2007}
J.-S. Park, J.~K. Jeong, Y.-G. Mo, H.~D. Kim, S.-I. Kim,
\newblock \emph{Applied Physics Letters} \textbf{2007}, \emph{90}, 26 262106.

\bibitem{Kim2012plasma}
J.~Kim, S.~Bang, S.~Lee, S.~Shin, J.~Park, H.~Seo, H.~Jeon,
\newblock \emph{Journal of Materials Research} \textbf{2012}, \emph{27}, 17
  2318.

\bibitem{Kim2014}
J.-S. Kim, M.-K. Joo, M.~X. Piao, S.-E. Ahn, Y.-H. Choi, H.-K. Jang, G.-T. Kim,
\newblock \emph{Journal of Applied Physics} \textbf{2014}, \emph{115}, 11
  114503.

\bibitem{Hwang2014}
Y.-H. Hwang, K.-S. Kim, W.-J. Cho,
\newblock \emph{Japanese Journal of Applied Physics} \textbf{2014}, \emph{53},
  4S 04EF12.

\bibitem{Liu2015}
X.~Liu, L.~L. Wang, H.~Hu, X.~Lu, K.~Wang, G.~Wang, S.~Zhang,
\newblock \emph{{IEEE} Electron Device Letters} \textbf{2015}, \emph{36}, 9
  911.

\bibitem{Lu2016}
H.~Lu, C.~Ren, X.~Xiao, Y.~Xiao, C.~Wang, S.~Zhang,
\newblock In \emph{2016 23rd International Workshop on Active-Matrix Flatpanel
  Displays and Devices ({AM}-{FPD})}. {IEEE}, \textbf{2016}
  \urlprefix\url{https://doi.org/10.1109/am-fpd.2016.7543642}.

\bibitem{Jang2018}
H.~Jang, S.~J. Lee, Y.~Porte, J.-M. Myoung,
\newblock \emph{Semiconductor Science and Technology} \textbf{2018}, \emph{33},
  3 035011.

\bibitem{Um2018}
J.~G. Um, J.~Jang,
\newblock \emph{Applied Physics Letters} \textbf{2018}, \emph{112}, 16 162104.

\bibitem{Liu2021}
W.-S. Liu, C.-H. Hsu, Y.~Jiang, Y.-C. Lai, H.-C. Kuo,
\newblock \emph{Semiconductor Science and Technology} \textbf{2021}, \emph{36},
  4 045007.

\bibitem{Park2022plasma}
H.~Park, J.~Yun, S.~Park, I.~sung Ahn, G.~Shin, S.~Seong, H.-J. Song, Y.~Chung,
\newblock \emph{{ACS} Applied Electronic Materials} \textbf{2022}.

\bibitem{Yasuta2021}
K.~Yasuta, T.~Ui, T.~Ikeda, D.~Matsuo, T.~Sakai, S.~Dohi, Y.~Setoguchi,
  E.~Takahashi, Y.~Andoh, J.~Tatemichi,
\newblock In \emph{2021 28th International Workshop on Active-Matrix Flatpanel
  Displays and Devices ({AM}-{FPD})}. {IEEE}, \textbf{2021}
  \urlprefix\url{https://doi.org/10.23919/am-fpd52126.2021.9499156}.

\bibitem{Ding2016}
X.~Ding, F.~Huang, S.~Li, J.~Zhang, X.~Jiang, Z.~Zhang,
\newblock \emph{Electronic Materials Letters} \textbf{2016}, \emph{13}, 1 45.

\bibitem{Abliz2020}
A.~Abliz,
\newblock \emph{Journal of Alloys and Compounds} \textbf{2020}, \emph{831}
  154694.

\bibitem{Joo2013}
Y.-H. Joo, J.-C. Woo, C.-I. Kim,
\newblock \emph{Microelectronic Engineering} \textbf{2013}, \emph{112} 74.

\bibitem{Wang2015}
G.~Wang, Z.~Song, X.~Xiao, S.~Zhang,
\newblock In \emph{2015 22nd International Workshop on Active-Matrix Flatpanel
  Displays and Devices ({AM}-{FPD})}. {IEEE}, \textbf{2015}
  \urlprefix\url{https://doi.org/10.1109%2Fam-fpd.2015.7173211}.

\bibitem{Kamiya2009}
T.~Kamiya, K.~Nomura, H.~Hosono,
\newblock \emph{Journal of Display Technology} \textbf{2009}, \emph{5}, 7 273.

\bibitem{Kim2012}
M.-S. Kim, Y.~H. Hwang, S.~Kim, Z.~Guo, D.-I. Moon, J.-M. Choi, M.-L. Seol,
  B.-S. Bae, Y.-K. Choi,
\newblock \emph{Applied Physics Letters} \textbf{2012}, \emph{101}, 24 243503.

\bibitem{Hsu2015}
C.-M. Hsu, W.-C. Tzou, C.-F. Yang, Y.-J. Liou,
\newblock \emph{Materials} \textbf{2015}, \emph{8}, 5 2769.

\bibitem{Jeong2016}
J.~Jeong, J.~Kim, D.~Kim, H.~Jeon, S.~M. Jeong, Y.~Hong,
\newblock \emph{{AIP} Advances} \textbf{2016}, \emph{6}, 8 085311.

\bibitem{Hu2017}
S.~Hu, K.~Lu, H.~Ning, Z.~Zheng, H.~Zhang, Z.~Fang, R.~Yao, M.~Xu, L.~Wang,
  L.~Lan, J.~Peng, X.~Lu,
\newblock \emph{{IEEE} Electron Device Letters} \textbf{2017}, \emph{38}, 7
  879.

\bibitem{Koretomo2020}
D.~Koretomo, S.~Hamada, Y.~Magari, M.~Furuta,
\newblock \emph{Materials} \textbf{2020}, \emph{13}, 8 1935.

\bibitem{Marrs2011}
M.~A. Marrs, C.~D. Moyer, E.~J. Bawolek, R.~J. Cordova, J.~Trujillo, G.~B.
  Raupp, B.~D. Vogt,
\newblock \emph{{IEEE} Transactions on Electron Devices} \textbf{2011},
  \emph{58}, 10 3428.

\bibitem{Jung2014}
H.~Y. Jung, Y.~Kang, A.~Y. Hwang, C.~K. Lee, S.~Han, D.-H. Kim, J.-U. Bae,
  W.-S. Shin, J.~K. Jeong,
\newblock \emph{Scientific Reports} \textbf{2014}, \emph{4}, 1.

\bibitem{Tai2019}
A.-H. Tai, C.-C. Yen, T.-L. Chen, C.-H. Chou, C.~W. Liu,
\newblock \emph{{IEEE} Transactions on Electron Devices} \textbf{2019},
  \emph{66}, 10 4188.

\bibitem{Cho2021}
M.~H. Cho, C.~H. Choi, H.~J. Seul, H.~C. Cho, J.~K. Jeong,
\newblock \emph{{ACS} Applied Materials {\&} Interfaces} \textbf{2021},
  \emph{13}, 14 16628.

\bibitem{Choi2022}
H.-W. Choi, K.-W. Song, S.-H. Kim, K.~T. Nguyen, S.~B. Eadi, H.-M. Kwon, H.-D.
  Lee,
\newblock \emph{Scientific Reports} \textbf{2022}, \emph{12}, 1.

\bibitem{Liu2017}
W.-S. Liu, Y.-H. Lin, C.-L. Huang, C.-W. Wang,
\newblock \emph{{IEEE} Transactions on Electron Devices} \textbf{2017},
  \emph{64}, 6 2533.

\bibitem{Dargar2019}
S.~K. Dargar, V.~M. Srivastava,
\newblock \emph{Micro {\&} Nano Letters} \textbf{2019}, \emph{14}, 13 1293.

\bibitem{Liu2022}
J.~Liu, S.~Liu, Y.~Yu, H.~Chen, C.~Wang, J.~Su, C.~Liu, Y.~Zhang, J.~Han,
  G.~Shao, Z.~Yao,
\newblock \emph{Advanced Electronic Materials} \textbf{2022}, 2100984.

\bibitem{Hsu2014}
H.-H. Hsu, C.-Y. Chang, C.-H. Cheng, S.-H. Chiou, C.-H. Huang,
\newblock \emph{{IEEE} Electron Device Letters} \textbf{2014}, \emph{35}, 1 87.

\bibitem{Stewart2017}
K.~A. Stewart, V.~Gouliouk, J.~M. McGlone, J.~F. Wager,
\newblock \emph{{IEEE} Transactions on Electron Devices} \textbf{2017},
  \emph{64}, 10 4131.

\bibitem{Lee2020}
H.~Lee, K.~Cho, D.~Kim, S.~Kim,
\newblock \emph{Semiconductor Science and Technology} \textbf{2020}, \emph{35},
  6 065014.

\bibitem{Kim2020}
D.~Kim, K.~Cho, S.~Woo, S.~Kim,
\newblock \emph{Electronics Letters} \textbf{2020}, \emph{56}, 2 102.

\bibitem{Kong2022}
H.~Kong, K.~Cho, H.~Lee, S.~Lee, J.~Lim, S.~Kim,
\newblock \emph{Materials Science in Semiconductor Processing} \textbf{2022},
  \emph{143} 106527.

\bibitem{Bang2017}
J.~Bang, S.~Matsuishi, H.~Hosono,
\newblock \emph{Applied Physics Letters} \textbf{2017}, \emph{110}, 23.

\bibitem{VanDeWalle2003}
C.~G.~V. de~Walle, J.~Neugebauer,
\newblock \emph{Nature} \textbf{2003}, \emph{423}, 6940 626.

\bibitem{VanDeWalle2006}
C.~G. {Van De Walle},
\newblock \emph{Physica B: Condensed Matter} \textbf{2006}, \emph{376-377}, 1
  1.

\bibitem{Tauc1968}
J.~Tauc,
\newblock \emph{Materials Research Bulletin} \textbf{1968}, \emph{3}, 1 37.

\bibitem{Park2022}
Y.-G. Park, D.~Y. Cho, R.~Kim, K.~H. Kim, J.~W. Lee, D.~H. Lee, S.~I. Jeong,
  N.~R. Ahn, W.-G. Lee, J.~B. Choi, M.~J. Kim, D.~Kim, S.~Jin, D.~G. Park,
  J.~Kim, S.~Choi, S.~Bang, J.~W. Lee,
\newblock \emph{Advanced Electronic Materials} \textbf{2022}, 2101273.

\bibitem{Li2013}
X.~Li, E.~Xin, L.~Chen, J.~Shi, J.~Zhang,
\newblock \emph{{AIP} Advances} \textbf{2013}, \emph{3}, 3 032137.

\bibitem{Luo2015}
D.~Luo, H.~Xu, M.~Zhao, M.~Li, M.~Xu, J.~Zou, H.~Tao, L.~Wang, J.~Peng,
\newblock \emph{{ACS} Applied Materials {\&} Interfaces} \textbf{2015},
  \emph{7}, 6 3633.

\bibitem{Ryu2012}
S.~H. Ryu, Y.~C. Park, M.~Mativenga, D.~H. Kang, J.~Jang,
\newblock \emph{{ECS} Solid State Letters} \textbf{2012}, \emph{1}, 2 Q17.

\bibitem{Nag2017}
M.~Nag, F.~D. Roose, K.~Myny, S.~Steudel, J.~Genoe, G.~Groeseneken,
  P.~Heremans,
\newblock \emph{Journal of the Society for Information Display} \textbf{2017},
  \emph{25}, 6 349.

\bibitem{Wager2010}
J.~F. Wager, D.~A. Keszler, R.~E. Presley,
\newblock \emph{Transparent electronics},
\newblock Springer, \textbf{2010}.

\end{thebibliography}
\bibliographystyle{MSP}
%\bibliography{MSP-template}
%\bibliographystyle{aipnum4-1}   %>>>> makes bibtex use Tspiebib.bst

\end{document}

\title{Supporting information for: Illuminating trap density trends in amorphous oxide semiconductors with ultrabroadband photoconduction}
\author{George W. Mattson$^1$, Kyle T. Vogt$^1$, John F. Wager$^2$ and Matt W. Graham$^1$}
\affiliation{Department of Physics, Oregon State University, Corvallis, OR 97331-6507, USA} 
\affiliation{School of EECS, Oregon State University, Corvallis, OR 97331-5501, USA}

\maketitle

\section{S.1 Average sub-gap DoS average trends between tog-gate(TG) and back-channel etch (BCE) a-IGZO TFTs } 

Supplement Figure S1 plots the density of states of six BCE a-IGZO TFTs ($\textit{color}$ circles) and two top-gate (TG) a-IGZO TFTs ($\textit{gray}$ squares). The mean DoS of the BCE a-IGZO TFTs (\textit{purple}) is shown against the mean DoS of the TG a-IGZO TFTs (\textit{black}). The BCE a-IGZO TFTs reflect a variety of different processing steps and/or characteristics compared to the as-grown BCE a-IGZO TFT (shown in \textit{blue}): Mo source-drain electrodes (Mo, \textit{red}) (as grown: Ti/Cu electrodes); reduced oxygen partial pressure during active channel deposition (80 \% $\mathrm{P_{O_2}}$, \textit{yellow}); low damage etch (low damage, \textit{green}); crystalline IGZO active channel composition (c-IGZO, \textit{orange}); and a wet-dry etch method (S/D dry+wet etch, \textit{yellow}). Regardless of the processing differences, the DoS of the BCE a-IGZO TFTs is universally higher in the range of -2.0 to -2.4 eV. The comparison of the DoS averages for the two different types of TFTs underlines the characteristically larger DoS of the BCE a-IGZO TFTs in this energetic region.
\begin{figure} [hbt]
   \begin{center}
   \begin{tabular}{c}
   \includegraphics[height=11cm]{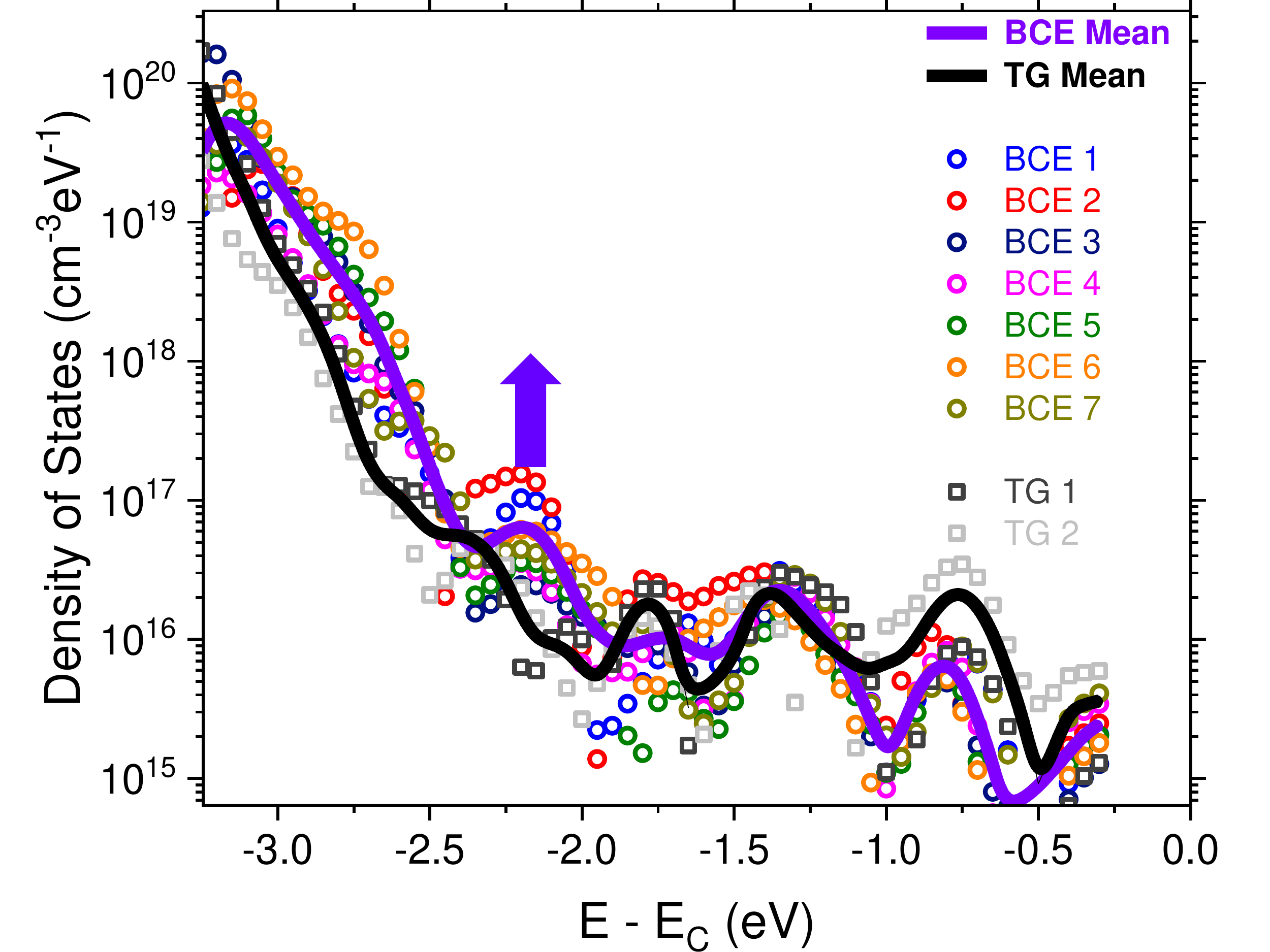}
   \end{tabular}
   \end{center}
   \caption[example] 
%>>>> use \label inside caption to get Fig. number with \ref{}
   { UBPC density of states spectra of six BCE a-IGZO TFTs (\textit{color}circles) and two top-gate a-IGZO TFTs (\textit{}black and gray squares). Purple and black curves show the mean TFT DoS of BCE devices relative to TG.  Purple arrow shows the peak at -2.2 eV for 7 different types of BCE TFTs consistent has more deep acceptor states present than all top-gate processing methods (where the enchant and TFT are never in contact).  } 
   \end{figure} 
   
   \section{S.2 TFT Etch Stop Layer (ESL) Impacts on Defect DoS } 

Supplement Figure S2 compares the total integrated trap density $\mathrm{N_{TOT}}$ of BCE a-IGZO TFTs with and without an etch stop layer intended to prevent the deleterious effects of etch-induced channel damage. The (\textit{Inset}) of Fig. S2 shows the drain current-gate voltage transfer curves of the two TFTs. The trap density spectrum of the TFT with the etch stop layer shows a large spike at $\mathrm{\sim -2.0}$ eV, but decreases between -2.0 and -2.7 eV, while the $\mathrm{N_{TOT}}$ of the BCE a-IGZO TFT does not show a large spike at -2.0 eV, and  gradually decreases between -2.0 and -2.7 eV. The increase in $\mathrm{N_{TOT}}$ over this range in the TFT without the ESL is consistent with the zinc vacancy peak centered at $\mathrm{\sim -2.2}$ eV, while the decrease observed in the spectrum of the ESL TFT over the same range is consistent with the lack of a zinc vacancy peak. The large difference in $\mathrm{N_{TOT}}$ between the two TFTs at $\mathrm{\sim -2.0}$ eV is curious. Is there a relationship between the suppression of traps at $\mathrm{\sim -2.0}$ eV and the emergence of traps at $\mathrm{\sim -2.2}$ eV? Further research into the energy of midgap defect formation under different processing conditions is motivated to address these questions.

\begin{figure}
   \begin{center}
   \begin{tabular}{c}
   \includegraphics[height=9cm]{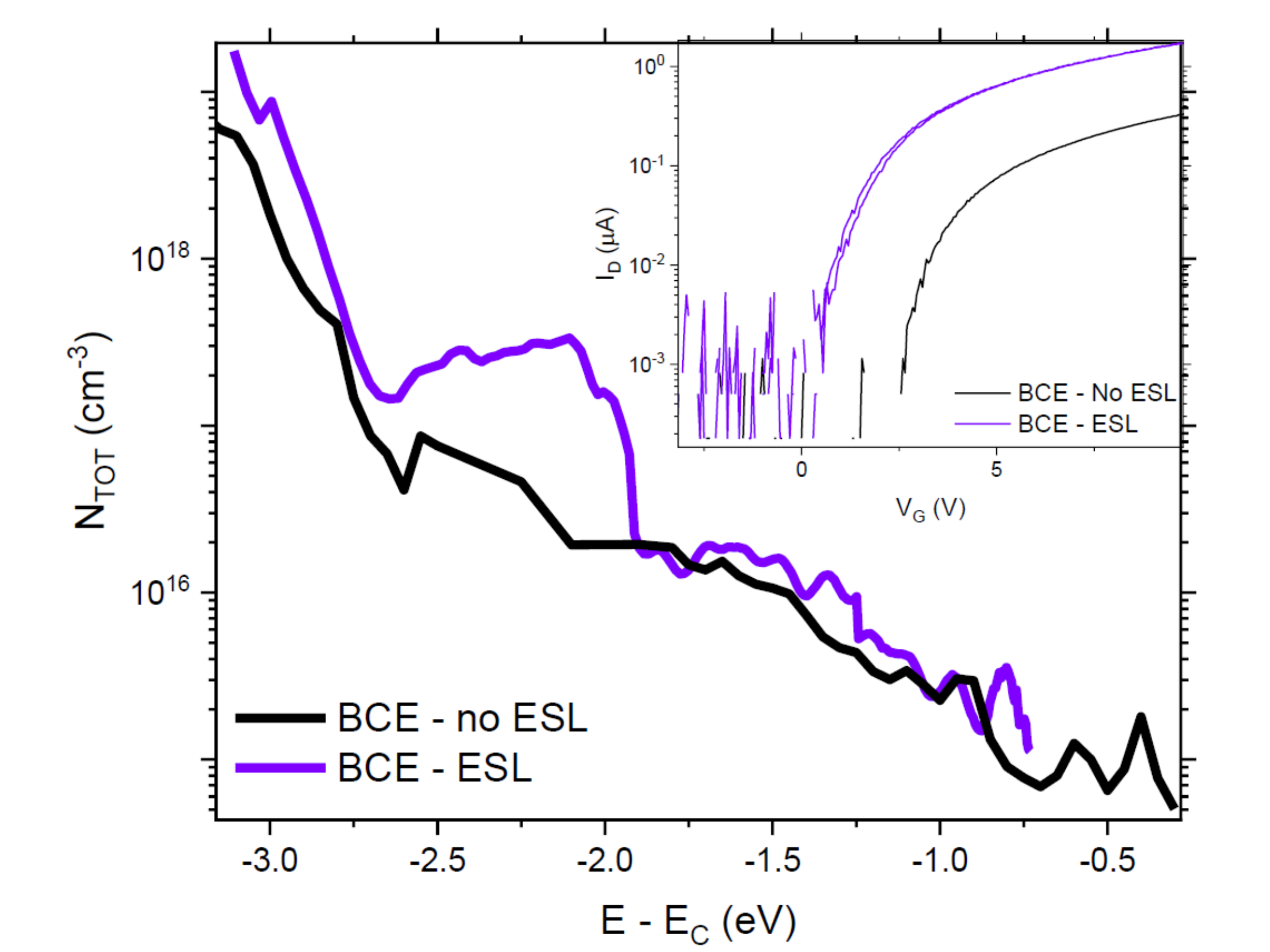}
   \end{tabular}
   \end{center}
   \caption[example] 
%>>>> use \label inside caption to get Fig. number with \ref{}
   { Total integrated trap density of BCE TFTs processed with and without an etch stop layer (ESL). (\textit{Inset}) Drain current-gate voltage transfer curves of BCE TFTs processed with and without an ESL.} 
   \end{figure} 
\newpage
   \section{S.3 Supplementary Experimental Detail on the UBPC DoS Experimental Apprartus}

\begin{figure}[ht]
   \begin{center}
   \begin{tabular}{c}
   \includegraphics[height=14cm]{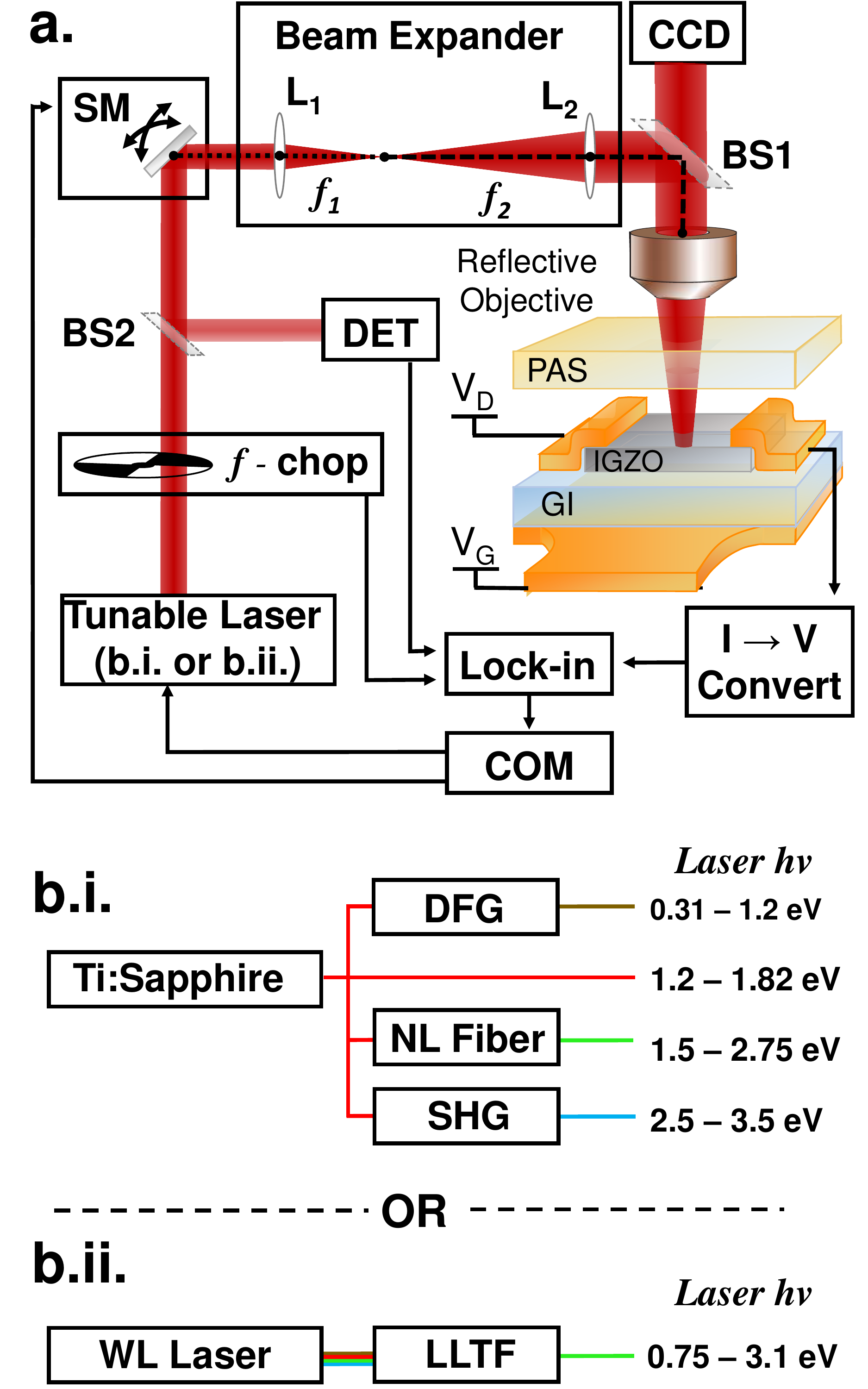}
   \end{tabular}
   \end{center}
   \caption[example] 
%>>>> use \label inside caption to get Fig. number with \ref{}
   { UBPC Experimental Setup. \textbf{(a)} Tunable lasers coupled into a scanning photocurrent microscope allow for continuous photoconduction measurement
over the range of the excitation source used. The scanning mirror (SM) and beam expanding lens pair allow for spatial control of the diffraction-limited laser spot, while a chopper modulates the beam so that the photoconduction can be recorded via lock-in detection. \textbf{(b)} Excitation sources are (i) the Ti:Sapphire-based system, which can cover laser energies from 0.31-3.5 eV, or (ii) the Supercontinuum White-Light (SCWL) laser coupled into a laser line tunable filter (LLTF) which acts as a monochromator.} 
   \end{figure}